\documentclass{article}
\usepackage[dvips]{graphics}
\newcommand{\mib}[1]{\mbox{\boldmath $#1$}}
\newcommand{\mibs}[1]{\mbox{\scriptsize\boldmath $#1$}}

\begin{document}
\title
{
\textbf{
Theory of Spin Fluctuation-Induced Superconductivity\\
Based on a $d$-$p$ Model\hspace*{5mm} II\\
|Superconducting State|}
}

\author
{ 
Tetsuya Takimoto
\footnote{Japan Science and Technology Corporation, 
Domestic Research Fellow.} and 
T$\rm{\hat o}$ru $\rm{Moriya}^{1}$ \\ 
\normalsize$Electrotechnical$  \normalsize$Laboratory,$  
\normalsize$Tsukuba,$  \normalsize$Ibaraki$  \normalsize$305$\\ 
\normalsize$^{1}Department$ \normalsize$of$ \normalsize$Physics,$ 
\normalsize$Faculty$ \normalsize$of$ 
\normalsize$Science$ \normalsize$and$ \normalsize$Technology, $
\normalsize$Science$ \normalsize$University$ 
\normalsize$of$ \normalsize$Tokyo,$ \normalsize$Noda$ \normalsize$278$
}

\date{(May \hspace{5mm}, 1998)}

\maketitle
\begin{abstract}
{The superconducting state of a two-dimensional $d$-$p$ model
is studied from the spin fluctuation point of view by using 
a strong coupling theory. 
The fluctuation exchange (FLEX) approximatoin is employed 
to calculate the spin fluctuations and 
the superconducting gap functions self-consistently 
in the optimal- and over-doped regions of hole concentration. 
The gap function has a symmetry of $d_{x^2 - y^2}$ type 
and develops below the transition temperature $T_{\rm c}$ 
more rapidly than in the BCS model. Its saturation value 
at the maximum is about $10 k_{\rm B}T_{\rm c}$. 
When the spin fluctuation-induced superconductivity is well 
stabilized at low temperatures in the optimal regime, 
the imaginary part of the antiferromagnetic spin susceptibility 
shows a very sharp resonance peak reminiscent of the 41 meV peak 
observed in the neutron scattering experiment on 
${\rm YBa}_{2}{\rm Cu}_{3}{\rm O}_{7}$. 
The one-particle spectral density around $\mib{k}=(\pi,0)$ 
shows sharp quasi-particle peaks followed by dip and hump 
structures bearing resemblance to the features observed in the 
angle-resolved photoemission experiment. 
With increasing doping concentration these features gradually 
disappear. }
\end{abstract}

\section{Introduction}
     Among a number of mechanisms proposed for 
the high temperature superconductivity since its discovery in 1986 
the spin fluctuation mechanism has continued to be one of 
the most promising mechanisms.
\cite{rf:1,rf:2,rf:3,rf:4,rf:5,rf:6,rf:7} 
Following systematic studies based on the parameterized 
spin fluctuation theory, more quantitative studies were carried out 
by using the Hubbard model
\cite{rf:8,rf:9,rf:10,rf:11,rf:12,rf:13,rf:14} 
and the $d$-$p$ model\cite{rf:15,rf:16,rf:17,rf:18} 
within the so-called fluctuation exchange (FLEX) approximation. 

     In a previous article\cite{rf:18} (hereafter referred to as I) 
we have developed a strong coupling theory of 
spin fluctuation-induced superconductivity on a microscopic basis 
by using a two-dimensional $d$-$p$ model and applied it to interpret 
high $T_{\rm c}$ cuprates. We adopted the fluctuation exchange 
(FLEX) approximation in calculating the transition temperature 
$T_{\rm c}$, the dynamical susceptibility, nuclear spin-lattice 
relaxation rate, and the one-particle spectral density above 
$T_{\rm c}$. The results compared rather well with 
the experimental results on cuprates in the optimal- and 
the over-doped regimes. 

     In the present article we extend the calculation to 
the superconducting state below $T_{\rm c}$. 
By using the $d$-$p$ model with the same parameter values as in I, 
we solve the non-linear Dyson-Gor'kov equations numerically and 
calculate various physical quantities; the energy gap function, 
the one-particle density of states, dynamical susceptibility 
and the one-particle spectral density. The results of calculation 
are discussed in comparison with the existing experimental results, 
particularly the resonance peak at 41 meV, observed in the 
neutron scattering experiment on 
${\rm YBa}_{2}{\rm Cu}_{3}{\rm O}_{7}$
\cite{rf:19,rf:20,rf:21,rf:22,rf:23} and the observed angle-resolved 
photoemission spectrum (ARPES), showing a resolution-limited 
quasi-particle peak followed by a dip and a hump structure 
around the $(\pi,0)$ point.\cite{rf:24,rf:25,rf:26,rf:27} 

     In what follows we first summarize the model and the formalism 
we use here in {\S} 2. In {\S} 3 the results of calculation are 
presented and discussed in comparison with the existing experimental 
results. Finally, {\S} 4 is devoted for summary of conclusions and 
general discussion.

\section{Model and Approach}

%

The model we use here is the same as in I; 
the $d$-$p$ model consisting of 
the $d_{x^2 - y^2}$ orbitals of copper atoms and 
the $p_{\sigma}$ ($p_{x}$ and $p_{y}$) orbitals of 
oxygen atoms in the ${\rm CuO_{2}}$ plane of cuprates. 
The energy levels of these orbitals are $\epsilon_{d}$ 
and $\epsilon_{p}$, respectively, and the transfer matrix 
elements are considered only between the neighboring $d$ and 
$p$ orbitals $(t_{dp})$ and between the neighboring $p_{x}$ and 
$p_{y}$ orbitals $(t_{pp})$ where the parities of both 
the $d_{x^2 - y^2}$ orbital and the $p_{\sigma}$ 
($p_{x}$ and $p_{y}$) orbitals are taken into account. 
For brevity, the Hubbard type electron-electron 
interaction $U$ is considered only within a $d$ orbital, and the 
other effects are assumed to be contained in $\epsilon_{d}$ 
and $\epsilon_{p}$ as the Hartree potentials. 
The Hamiltonian may be written as follows:
\begin{eqnarray}
  H&=&H_{\rm 0}+H_{\rm I} ,\\
  H_{\rm 0}&=&\sum_{{\mibs{k}},\sigma}
  [\epsilon_{d}d_{{\mibs{k}}\sigma}^{\dagger}d_{{\mibs{k}}\sigma}
  +\sum_{m=1}^{2}[\{\epsilon_{p}-(-1)^{m}t_{p}(\mib k)\}
  p_{m{\mibs{k}}\sigma}^{\dagger}p_{m{\mibs{k}}\sigma} 
  +\{{\rm i}t_{m}(\mib k)d_{{\mibs{k}}\sigma}^{\dagger}
   p_{m{\mibs{k}}\sigma}
  +h.c.\}]] , \mib{\mbox{}}
\end{eqnarray}
with
\begin{eqnarray}
  &&t_{m}({\mib k})=\sqrt{2}t_{dp}
  [\sin(k_{x}a/2)-(-1)^{m}\sin(k_{y}a/2)],\nonumber\\
  &&t_{p}({\mib k})=4t_{pp}
  \sin(k_{x}a/2)\sin(k_{y}a/2),
\end{eqnarray}
and
\begin{eqnarray}
  H_{\rm I}&=&U\sum_{j}n_{dj\uparrow}n_{dj\downarrow}\nonumber\\
  &=&\frac{U}{N_{0}}\sum_{\mibs q}\sum_{\mibs k}\sum_{\mibs k'}
  d_{\mibs k+\mibs q\uparrow}^{\dagger}
  d_{\mibs k'-\mibs q\downarrow}^{\dagger}
  d_{\mibs k'\downarrow}d_{\mibs k\uparrow}, \mib{\mbox{}}
\end{eqnarray}
where $d_{\mibs{k}\sigma}\mib{\mbox{}}$ and 
$p_{m\mibs{k}\sigma}\mib{\mbox{}}$($m=1,2$) are 
the Fourier transforms of the annihilation operators for the 
electrons in the $d$ and 
$p_{m}=\frac{1}{\sqrt{2}}[p_{x}-(-1)^{m}p_{y}]$ 
orbitals, respectively, and $N_{\rm 0}$ is the number of atoms 
in the crystal.

We denote the diagonalized expression for $H_{\rm 0}$ as follows:
\begin{equation}
  H_{\rm 0}=\sum_{\mibs k,\sigma}\sum_{m=1}^{3}
  E_{m\mibs k\sigma}
  a_{m\mibs k\sigma}^{\dagger}a_{m\mibs k\sigma}, 
  \mib{\mbox{}}\nonumber
\end{equation}\vspace*{-2mm}
\begin{eqnarray}
  &&\left[\begin{array}{c}a_{1\mibs k\sigma}\mib{\mbox{}}\\
                        a_{2\mibs k\sigma}\mib{\mbox{}}\\
                        a_{3\mibs k\sigma}\mib{\mbox{}}
        \end{array}\right]
 =\left[\begin{array}{ccc}
   \beta_{11\mibs k}\mib{\mbox{}} & \beta_{12\mibs k}\mib{\mbox{}} & 
   \beta_{13\mibs k} \mib{\mbox{}}\\
   \beta_{21\mibs k}\mib{\mbox{}} & \beta_{22\mibs k}\mib{\mbox{}} & 
   \beta_{23\mibs k} \mib{\mbox{}}\\
   \beta_{31\mibs k}\mib{\mbox{}} & \beta_{32\mibs k}\mib{\mbox{}} & 
   \beta_{33\mibs k}\mib{\mbox{}}
        \end{array}\right]
  \left[\begin{array}{c}d_{\mibs k\sigma}\mib{\mbox{}}\\
                        p_{1\mibs k\sigma}\mib{\mbox{}}\\
                        p_{2\mibs k\sigma}\mib{\mbox{}}
        \end{array}\right].\mib{\mbox{}}
\end{eqnarray}
     Then the Green's functions for the non-interacting system is 
given by
\begin{equation}
  G_{\lambda\nu}^{(0)}({\mib k},{\rm i}\omega_{n})
  =\sum_{m=1}^{3}\beta_{m\lambda\mibs k}^{\ast}\beta_{m\nu\mibs k}
  \frac{1}{{\rm i}\omega_{n}+\mu-E_{m\mibs k\sigma}} ,\mib{\mbox{}}
\end{equation}
where $\lambda(\nu)$=1,2 and 3 stand for $d$, $p_{1}$ and $p_{2}$ 
orbitals, respectively, $\omega_{n}=(2n+1)\pi T$ is the 
Fermion Matsubara frequency and $\mu$ is the chemical potential. 
$E_{m\mibs k\sigma}\mib{\mbox{}}$ and 
$\beta_{m\lambda\mibs k}\mib{\mbox{}}$ can be calculated 
from $\epsilon_{d}$, $\epsilon_{p}$, $t_{dp}$ and $t_{pp}$. 
In what follows we express all the energies in units of $t_{dp}$. 

     The Dyson-Gor'kov equations for the Green's functions
$G_{\lambda\nu}({\mib k},{\rm i}\omega_{n})$ and the anormalous 
Green's functions $F_{\lambda\nu}^{\dagger}
({\mib k},{\rm i}\omega_{n})$ are given by
\begin{eqnarray}
  &&G_{\lambda\nu}({\mib k},{\rm i}\omega_{n})
  =G_{\lambda\nu}^{(0)}({\mib k},{\rm i}\omega_{n})
  +G_{\lambda 1}^{(0)}({\mib k},{\rm i}\omega_{n})
  [\Sigma^{(1)}(\mib k,{\rm i}\omega_{n})
  G_{1\nu}({\mib k},{\rm i}\omega_{n})
  -\Sigma^{(2)}(\mib k,{\rm i}\omega_{n})
  F_{1\nu}^{\dagger}({\mib k},{\rm i}\omega_{n})],\\
  \nonumber\\
  &&\vspace*{-2mm}F_{\lambda\nu}^{\dagger}({\mib k},{\rm i}\omega_{n})
  =G_{\lambda 1}^{(0)}({-\mib k},-{\rm i}\omega_{n})
  [\Sigma^{(1)}(-\mib k,-{\rm i}\omega_{n})
  F_{1\nu}^{\dagger}({\mib k},{\rm i}\omega_{n})
  +\Sigma^{(2)}(-\mib k,-{\rm i}\omega_{n})
  G_{1\nu}({\mib k},{\rm i}\omega_{n})],
\end{eqnarray}
where the self-energies due to the spin and charge fluctuations 
are given within the FLEX approximation as follows:
\begin{eqnarray}
  \Sigma^{(1)}(\mib k,{\rm i}\omega_{n})
  &=&\frac{T}{N_{\rm 0}}\sum_{\mibs q,m}
  V_{\rm eff}(\mib q,{\rm i}\Omega_{m})
  G_{dd}({\mib k-\mib q},{\rm i}\omega_{n}-{\rm i}\Omega_{m}),\\
  \Sigma^{(2)}(\mib k,{\rm i}\omega_{n})
  &=&-\frac{T}{N_{\rm 0}}\sum_{\mibs q,m}
  V_{\rm sing}(\mib q,{\rm i}\Omega_{m})
  F_{dd}({\mib k-\mib q},{\rm i}\omega_{n}-{\rm i}\Omega_{m}),
\end{eqnarray}
with
\begin{eqnarray}
  &&V_{\rm eff}(\mib q,{\rm i}\Omega_{m})
  =U+U^{2}[\frac{3}{2}\chi_{d}^{\rm s}(\mib q,{\rm i}\Omega_{m})
  +\frac{1}{2}\chi_{d}^{\rm c}(\mib q,{\rm i}\Omega_{m})
  -\frac{1}{2}\{
  \overline{\chi}_{d}^{\rm s}(\mib q,{\rm i}\Omega_{m})
  +\overline{\chi}_{d}^{\rm c}(\mib q,{\rm i}\Omega_{m})\}],\\
  &&V_{\rm sing}(\mib q,{\rm i}\Omega_{m})
  =U+U^{2}[\frac{3}{2}\chi_{d}^{\rm s}(\mib q,{\rm i}\Omega_{m})
  -\frac{1}{2}\chi_{d}^{\rm c}(\mib q,{\rm i}\Omega_{m})
  -\frac{1}{2}\{
  \overline{\chi}_{d}^{\rm s}(\mib q,{\rm i}\Omega_{m})
  -\overline{\chi}_{d}^{\rm c}(\mib q,{\rm i}\Omega_{m})\}],
\end{eqnarray}
and
\begin{eqnarray}
  &&\chi_{d}^{\rm s}(\mib q,{\rm i}\Omega_{m})
  =\frac{\overline{\chi}_{d}^{\rm s}(\mib q,{\rm i}\Omega_{m})}
  {1-U\overline{\chi}_{d}^{\rm s}(\mib q,{\rm i}\Omega_{m})}
  ,\hspace{1cm}
  \chi_{d}^{\rm c}(\mib q,{\rm i}\Omega_{m})
  =\frac{\overline{\chi}_{d}^{\rm c}(\mib q,{\rm i}\Omega_{m})}
  {1+U\overline{\chi}_{d}^{\rm c}(\mib q,{\rm i}\Omega_{m})},\\
  &&\nonumber\\
  &&\overline{\chi}_{d}^{\rm s}(\mib q,{\rm i}\Omega_{m})
  =-\frac{T}{N_{0}}\sum_{\mibs k,n}
  [G_{dd}({\mib k+\mib q},{\rm i}\omega_{n}+{\rm i}\Omega_{m})
   G_{dd}({\mib k},{\rm i}\omega_{n})
  +F_{dd}({\mib k+\mib q},{\rm i}\omega_{n}+{\rm i}\Omega_{m})
   F_{dd}({\mib k},{\rm i}\omega_{n})],\\
  &&\overline{\chi}_{d}^{\rm c}(\mib q,{\rm i}\Omega_{m})
  =-\frac{T}{N_{0}}\sum_{\mibs k,n}
  [G_{dd}({\mib k+\mib q},{\rm i}\omega_{n}+{\rm i}\Omega_{m})
   G_{dd}({\mib k},{\rm i}\omega_{n})
  -F_{dd}({\mib k+\mib q},{\rm i}\omega_{n}+{\rm i}\Omega_{m})
   F_{dd}({\mib k},{\rm i}\omega_{n})],
\end{eqnarray}
where $\Omega_{m}=2m\pi T$ is the Bose Matsubara frequency. 

     The number of $d$- and $p$-electrons 
per site (respectively, $n_{d}$ and $n_{p}$) and 
the doping concentration $\delta$ are given by
\begin{eqnarray}
  &&n_{d}=2\lim_{\tau\rightarrow +0}\frac{T}{N_{\rm 0}}
  \sum_{\mibs k,n}{\rm e}^{{\rm i}{\omega_{n}}{\tau}}
  G_{dd}({\mib k},{\rm i}\omega_{n}),\nonumber\\
  &&n_{p}=\lim_{\tau\rightarrow +0}\frac{T}{N_{\rm 0}}
  \sum_{\mibs k,n}{\rm e}^{{\rm i}\omega_{n}{\tau}}
  [G_{22}({\mib k},{\rm i}\omega_{n})
  +G_{33}({\mib k},{\rm i}\omega_{n})],\nonumber\\
  &&\nonumber\\
  &&\delta=5-(n_{d}+2n_{p}),\nonumber
\end{eqnarray}
where the "half-filled" state corresponds to 
$n_{d}+2n_{p}=5$. 

     For the purpose of calculating $T_{\rm c}$ we linearize 
eqs.(8) and (9) with respect
to $F_{dd}^{\dagger}({\mib k},{\rm i}\omega_{n})$ or 
$\Sigma^{(2)}(\mib k,{\rm i}\omega_{n})$ as follows:
\begin{eqnarray}
  &&\frac{1}{G_{dd}({\mib k},{\rm i}\omega_{n})}
  =\frac{1}{G_{dd}^{(0)}({\mib k},{\rm i}\omega_{n})}
  -\Sigma^{(1)}(\mib k,{\rm i}\omega_{n}),\\
  &&\nonumber\\
  &&F_{dd}^{\dagger}({\mib k},{\rm i}\omega_{n})
  =G_{dd}({-\mib k},-{\rm i}\omega_{n})
  \Sigma^{(2)}(-\mib k,-{\rm i}\omega_{n})
  G_{dd}({\mib k},{\rm i}\omega_{n}).
\end{eqnarray}
The normal Green's functions are now calculated self-consistently 
from eqs.(10), (12), (14) and (17). The critical temperature for 
the superconductivity is determined as the temperature below 
which the linearized equation for 
$\Sigma^{(2)}(\mib k,{\rm i}\omega_{n})$ has a non-trivial solution.
The linearized equation is given by
\begin{eqnarray}
  \Sigma^{(2)}(\mib k,{\rm i}\omega_{n})
  =-\frac{T}{N_{\rm 0}}\sum_{\mibs p\mib{\mbox{}},m}
  [V_{\rm sing}(\mib k-\mib p,{\rm i}\omega_{n}-{\rm i}\omega_{m})
  |G_{dd}({\mib p},{\rm i}\omega_{m})|^{2}]_{F_{dd}\rightarrow 0}
  \Sigma^{(2)}(\mib p,{\rm i}\omega_{m}).
\end{eqnarray}
$T_{\rm c}$ is thus obtained by solving this eigenvalue problem, 
eq.(19); $T_{\rm c}$ is the temperature where the maximum eigenvalue 
becomes unity. The superconducting order parameter has the same 
symmetry as $\Sigma^{(2)}(\mib k,{\rm i}\omega_{n})$.

     Below $T_{\rm c}$ we need to solve the nonlinear equations 
(from (7) to (16)) self-consistently. From the solution we calculate 
the one-particle spectra $A({\mib k},\omega)$ 
and the transverse dynamical spin susceptibility 
$\chi^{-+}(\mib q,\omega)$ by analytically continuing 
$G_{dd}({\mib k},{\rm i}\omega_{n})$ 
and $\chi_{d}^{\rm s}(\mib q,{\rm i}\Omega_{m})$ 
from the imaginary-axis to the real-axis. 
The density of states $\rho(\omega)$ and 
the local spin susceptibility $\chi(\omega)$ are given as follows:
\begin{eqnarray}
  &&\rho(\omega)=\frac{1}{N_{0}}\sum_{\mibs k}
  A({\mib k},\omega), \\
  &&A({\mib k},\omega)
  =-\frac{1}{\pi}{\rm Im}
  [G_{dd}({\mib k},\omega+{\rm i}\eta)
  +G_{22}({\mib k},\omega+{\rm i}\eta)
  +G_{33}({\mib k},\omega+{\rm i}\eta)], \\
  &&{\rm Im}\chi(\omega)
  =\frac{1}{N_{0}}\sum_{\mibs q}
  {\rm Im}\chi^{-+}(\mib q,\omega), \hspace*{10 mm}
  \chi^{-+}(\mib q,\omega)
  =\chi_{d}^{\rm s}(\mib q,\omega+{\rm i}\eta), \hspace*{10 mm}
  (\eta\rightarrow +0).
\end{eqnarray}

\begin{table}[t]
\caption{Parameter values for the $d$-$p$ model and 
the calculated transition temperature $T_{\rm c}$ in units of 
$t_{dp}$ 
for various doping concentrations.}
\label{table:1}
\begin{tabular}{p{16.3mm}p{30.3mm}p{16.3mm}
p{30.3mm}p{16.3mm}p{30.3mm}} \hline
\centering $\epsilon_{p}-\mu$ &\centering $n_d$  &\centering $n_p$  &\centering  $(2-n_d)/(4-2n_p)$  &\centering
$\delta$  &\multicolumn{1}{c}{$T_{\rm c}$}\\ \hline
\centering -2.600 &\centering 1.314 &\centering 1.799 &\centering 
1.703 &\centering 0.0890 &\hfill 0.006621\hfill\null\\
\centering -2.575 &\centering 1.308 &\centering 1.794 &\centering 
1.683 &\centering 0.1029 &\hfill 0.006597\hfill\null\\
\centering -2.550 &\centering 1.302 &\centering 1.790 &\centering 
1.663 &\centering 0.1170 &\hfill 0.006475\hfill\null\\
\centering -2.500 &\centering 1.291 &\centering 1.782 &\centering 
1.624 &\centering 0.1451 &\hfill 0.006015\hfill\null\\
\centering -2.450 &\centering 1.280 &\centering 1.773 &\centering 
1.585 &\centering 0.1738 &\hfill 0.005351\hfill\null\\
\centering -2.400 &\centering 1.269 &\centering 1.764 &\centering 
1.547 &\centering 0.2037 &\hfill 0.004529\hfill\null\\ \hline
\end{tabular}
\end{table}

\begin{figure}[ht]
\begin{center}
\resizebox{70mm}{!}{\includegraphics{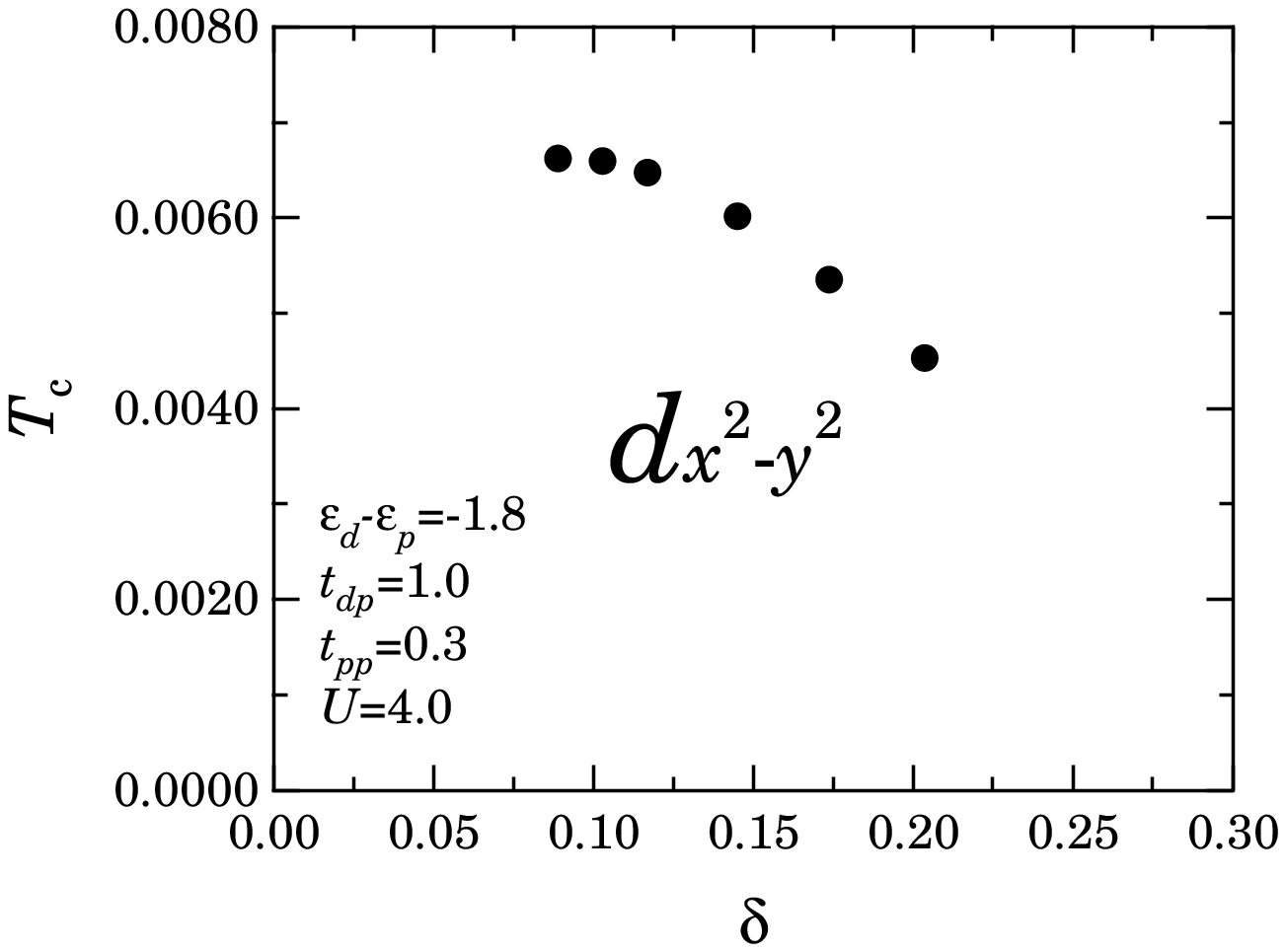}}
\end{center}
{\small 
\begin{center}
Fig. 1. 
Calculated values of $T_{\rm c}$ for various 
doping concentrations.
\end{center}}
\end{figure}

We define the gap functions in the following conventional way:
\begin{eqnarray}
  &&\Delta({\mib k},\omega)=\frac
  {\Sigma^{(2)}(\mib k,\omega+{\rm i}\eta)}
  {Z(\mib k,\omega)},\\
  &&\omega Z(\mib k,\omega)
  =\omega-\frac{1}{2}[\Sigma^{(1)}(\mib k,\omega+{\rm i}\eta)
  -\Sigma^{(1)}(\mib k,-\omega-{\rm i}\eta)].
\end{eqnarray}

\section{Results of Calculation}

\subsection{Some details of numerical calculations.}
     To study the superconducting state, 
we first calculate the superconducting transition temperature 
$T_{\rm c}$. For this purpose we calculate the kernel of 
the eigenvalue equation (19) at a fixed temperature, 
solving eqs. (10) and (12-17) for 
$\chi_{d}^{\rm s}(\mib q,{\rm i}\Omega_{m})$, 
$\chi_{d}^{\rm c}(\mib q,{\rm i}\Omega_{m})$ 
and $G_{dd}({\mib k},{\rm i}\omega_{n})$ self-consistently. 
Then, using the results of this calculation, 
we solve the eigenvalue problem. If eq. (19) do not have 
a non-trivial solution at a given temperature, we carry out 
the same calculation at a slightly decreased temperature. 
We repeat this routine until a error for $T_{\rm c}$ 
attains $10^{-6}$. Below $T_{\rm c}$, we calcurate 
$\chi_{d}^{\rm s}(\mib q,{\rm i}\Omega_{m})$, 
$\chi_{d}^{\rm c}(\mib q,{\rm i}\Omega_{m})$, 
$G_{dd}({\mib k},{\rm i}\omega_{n})$ and 
$F_{dd}({\mib k},{\rm i}\omega_{n})$ by solving the non-linear 
equations (7) to (16), self-consistently. 

All the sums involved in the self-consistent equations are carried 
out by using fast Fourier transforms (FFT) with the $64\times64$ 
meshes in the $\mib k$-space and the Matsubara frequency sum up to 
$|\omega_{n}|\approx25|t_{dp}|$. 
The solution is obtained by iteration until the self-consistency 
condition is satisfied within the following accuracy:
\begin{eqnarray}
  \frac{|\Sigma^{(m)}_{r}(\mib k,{\rm i}\omega_{n})
  -\Sigma^{(m)}_{r-1}(\mib k,{\rm i}\omega_{n})|}
  {|\Sigma^{(m)}_{r}(\mib k,{\rm i}\omega_{n})|}<10^{-6}
  \hspace*{5mm}m=1, 2.
\end{eqnarray}
The analytical continuations to the real frequency axis are 
carried out by using Pad\'{e} approximants. 

We need to take special care in evaluating 
$\rho(\omega)$ and ${\rm Im}\chi(\omega)$ 
because of the size-effect due to the striking reduction 
of the quasi-particle damping in the superconducting state; 
after substituting the $\mib k$-summations for 
$G_{dd}({\mib k},{\rm i}\omega_{n})$ or 
$\chi_{d}^{\rm s}(\mib q,{\rm i}\Omega_{m})$ with numerical 
integrations by using the spline interpolation method, 
the Pad\'{e} approximants are calculated. 

Now, for the actual calculations, 
we choose the same parameter values as in I, 
$t_{pp}=0.3, \epsilon_{d}-\epsilon_{p}=-1.8, U=4$, 
in units of $t_{dp}$.

\subsection{Transition temperature.}
     Although the present calculation has a better accuracy than 
the previous one, the results as shown in Table I and Fig. 1 
are only slightly different from those in I. 
The calculated value of $T_{\rm c}$ and its doping concentration 
dependence are reasonable in the optimal and over-doped regimes 
while the observed reduction of $T_{\rm c}$ with decreasing 
doping concentration in the under-doped regime is hard to reproduce. 
The pairing symmetry of the calculated state is always of 
$d_{x^2 - y^2}$. Thus it seems appropriate to confine ourselves 
in what follows to the superconducting state of this symmetry.

\begin{figure}[ht]
\begin{center}
\resizebox{70mm}{!}{\includegraphics{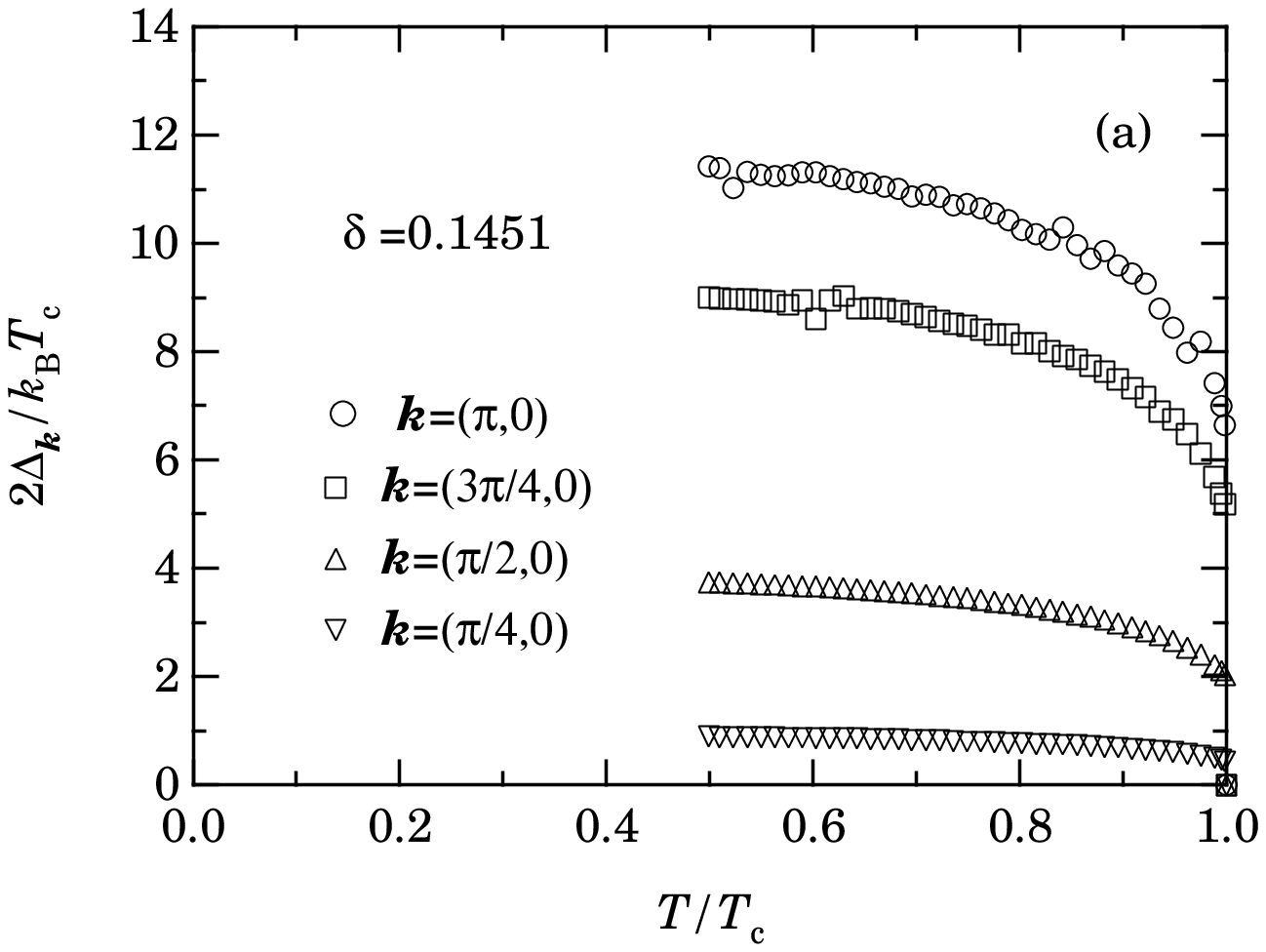}}\hspace*{5mm}
\resizebox{70mm}{!}{\includegraphics{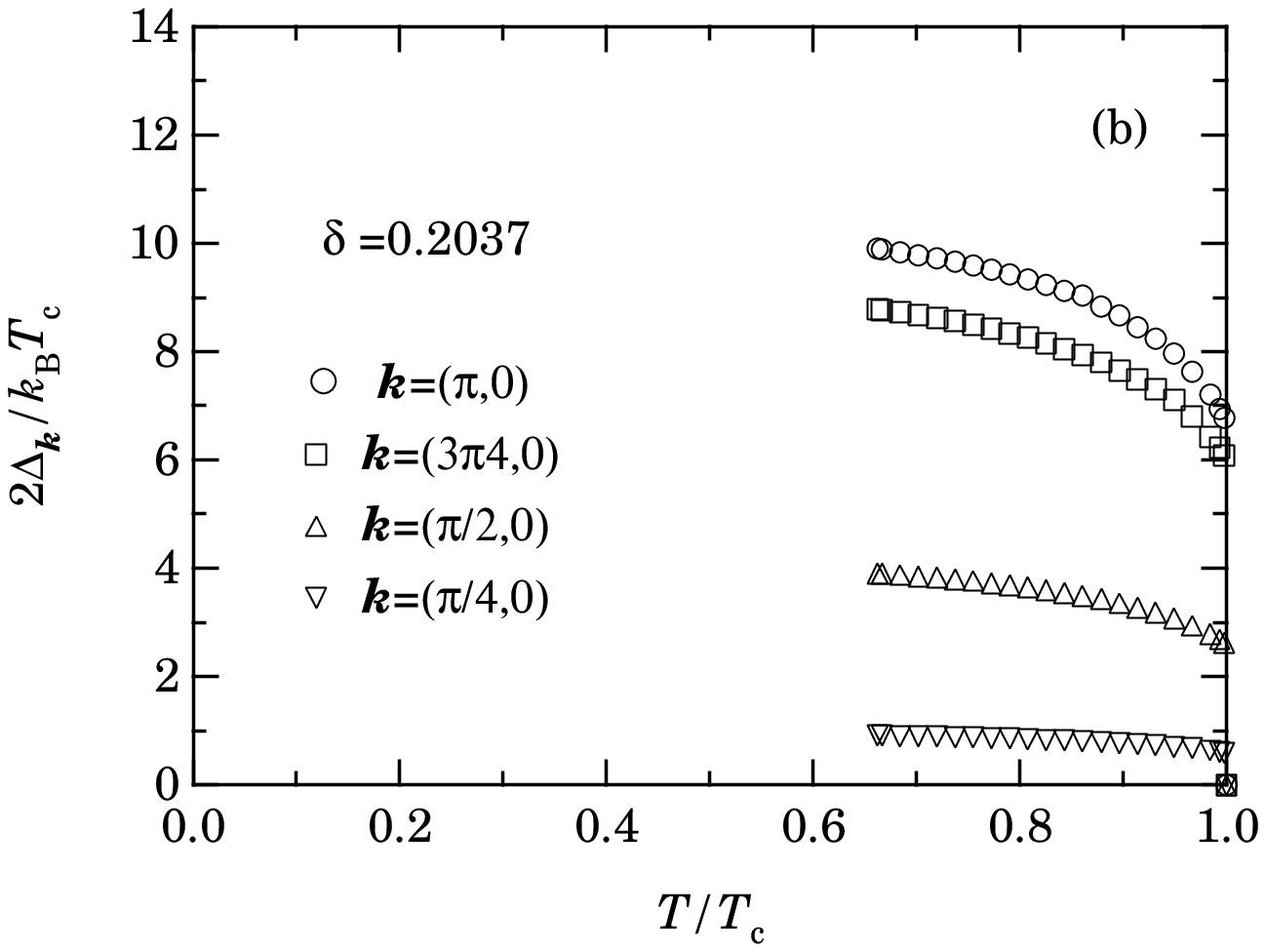}}
\end{center}
{\small 
\begin{flushleft}
Fig. 2. 
Temperature dependence of 
$2\Delta_{\mibs k\mib{\mbox{}}}/k_{\rm B}T_{\rm c}$ 
for $\mib{k}=(\pi,0)$, $(3\pi/4,0)$, $(\pi/2,0)$ and $(\pi/4,0)$. 
(a) $\delta=0.1451$, (b) $\delta=0.2037$.
\end{flushleft}}
\end{figure}

\begin{figure}[ht]
\begin{center}
\resizebox{70mm}{!}{\includegraphics{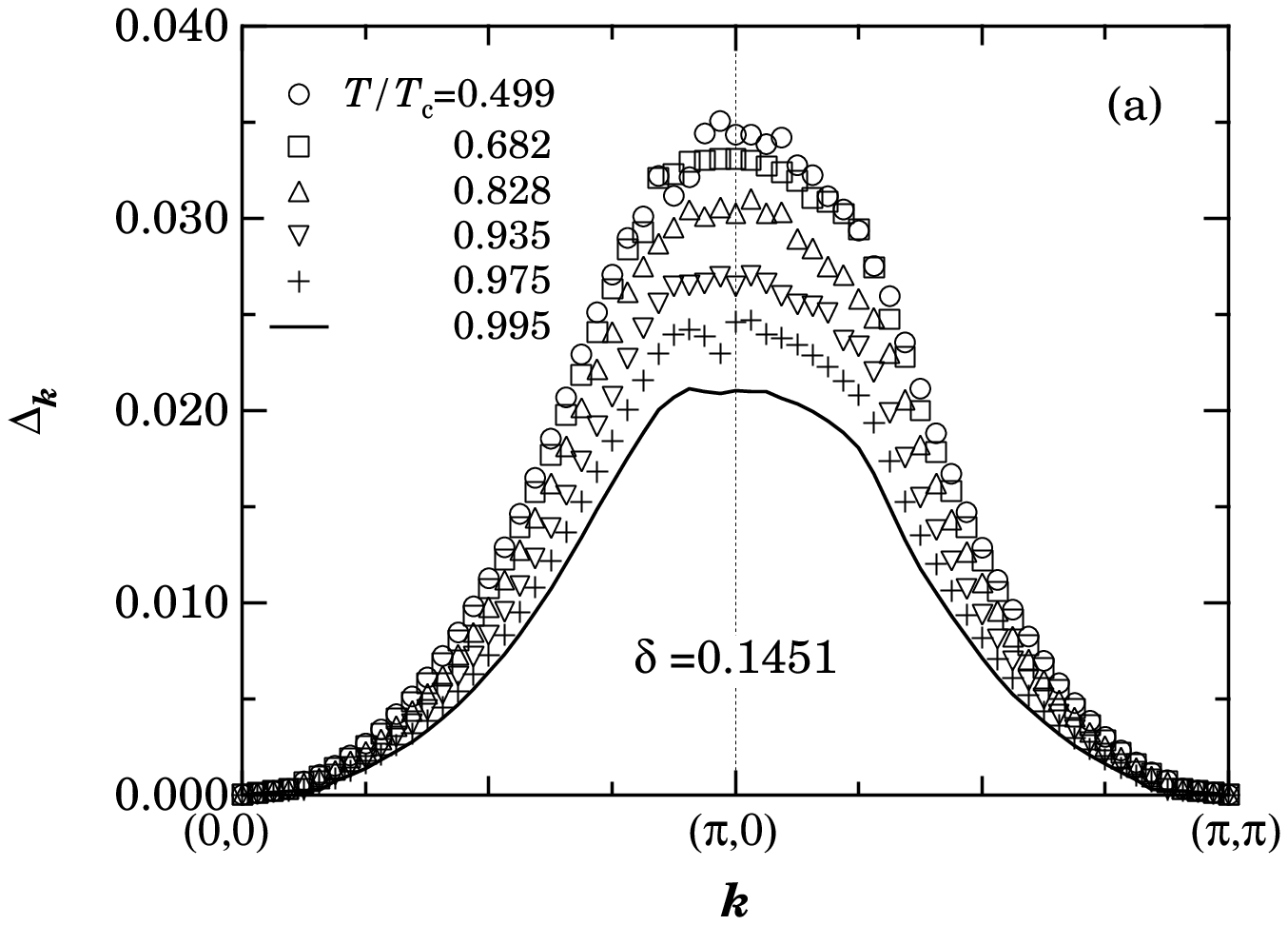}}\hspace*{5mm}
\resizebox{70mm}{!}{\includegraphics{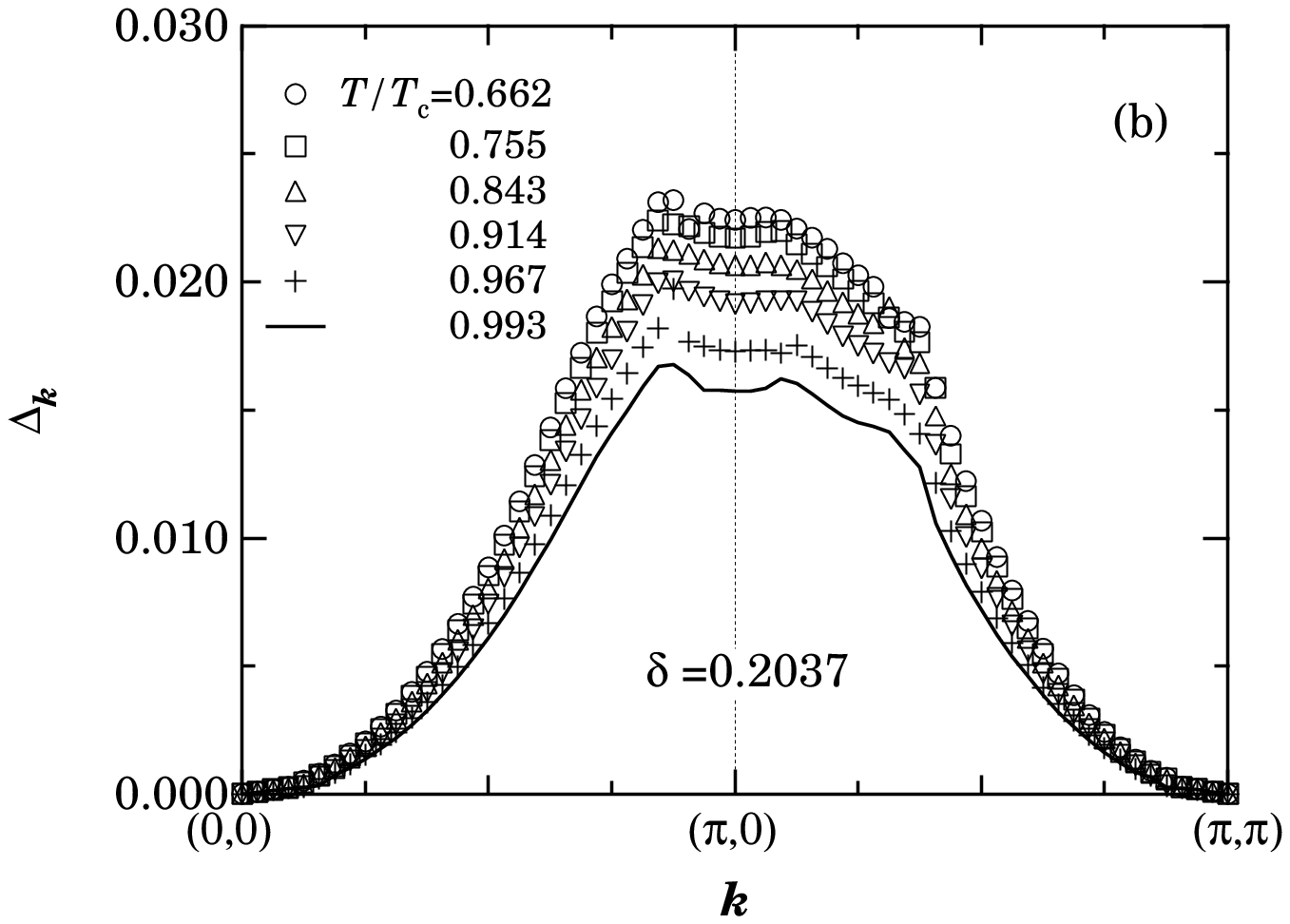}}
\end{center}
{\small 
\begin{center}
Fig. 3. 
Anisotropy of the gap function at various temperatures. 
(a) $\delta=0.1451$, (b) $\delta=0.2037$.
\end{center}}
\end{figure}

\subsection{Gap function.}
     We calculate the gap function as defined conventionally 
in the $\acute{\rm E}$liashberg theory; 
$\Delta_{\mibs k\mib{\mbox{}}}=\Delta({\mib k},
\Delta_{\mibs k\mib{\mbox{}}})$. Fig. 2(a) shows the temperature 
dependence of $2\Delta_{\mibs k\mib{\mbox{}}}/k_{\rm B}T_{\rm c}$ 
for $\mib{k}=(\pi,0)$, $(3\pi/4,0)$, $(\pi/2,0)$ and $(\pi/4,0)$ 
for $\delta=0.1451$ and Fig. 2(b) shows the corresponding results 
for $\delta=0.2037$. We find that the superconducting gap 
develops more rapidly than in the BCS model. The maximum gap at 
$\mib{k}=(\pi,0)$ saturates to a low temperature value of around 
$10k_{\rm B}T_{\rm c}$. We next show in Figs. 3(a) and 3(b) 
the $\mib k$-dependence of $\Delta_{\mibs k\mib{\mbox{}}}$ at 
varying temperatures for $\delta=0.1451$ and $\delta=0.2037$, 
respectively. The results are roughly proportional to 
$|\cos{k_{x}}-\cos{k_{y}}|$.

\begin{figure}[t]
\begin{center}
\resizebox{70mm}{!}{\includegraphics{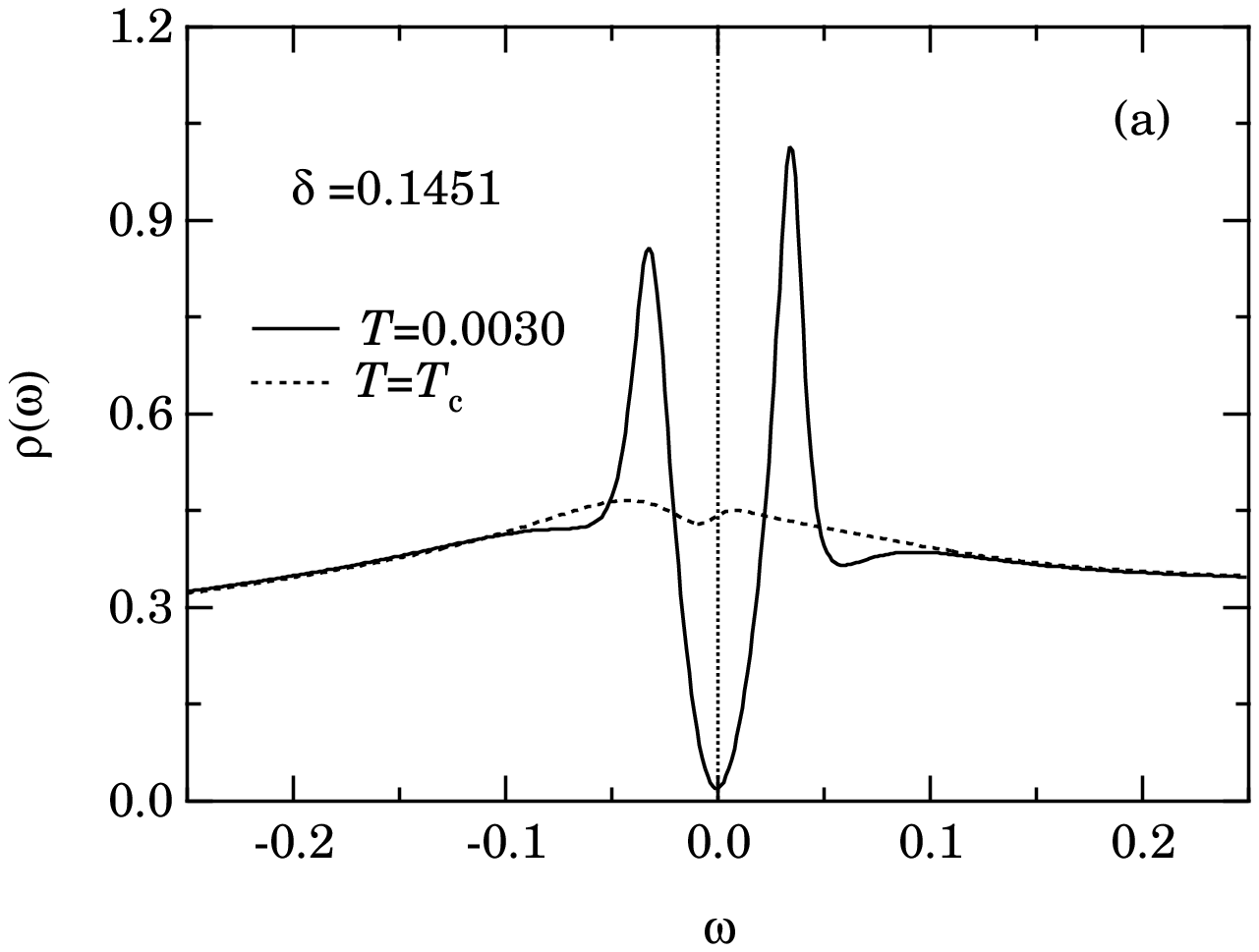}}\hspace*{5mm}
\resizebox{70mm}{!}{\includegraphics{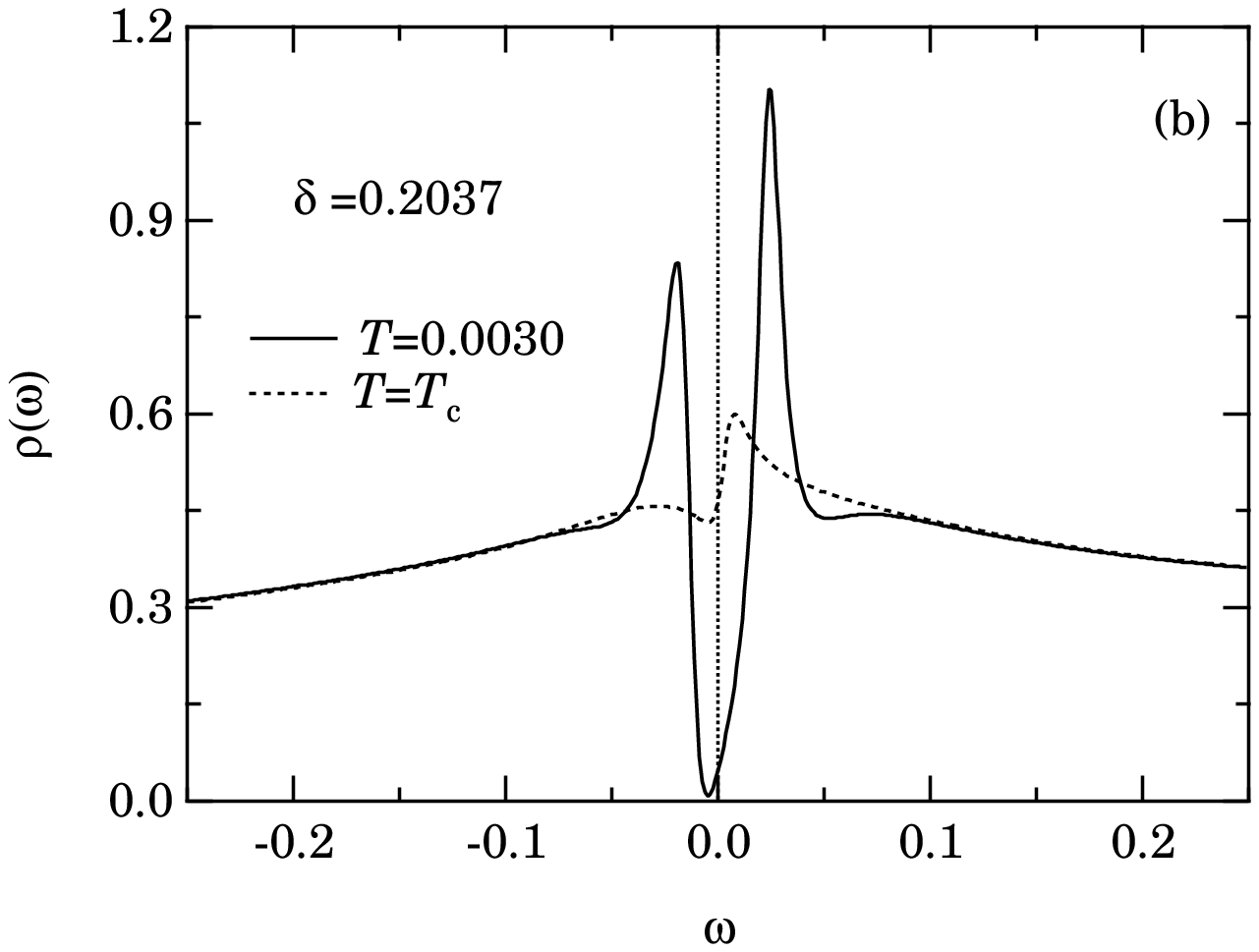}}
\end{center}
{\small 
\begin{center}
Fig. 4. 
One particle density of states. 
(a) $\delta=0.1451$, (b) $\delta=0.2037$.
\end{center}}
\end{figure}

\subsection{Density of states.}
     Figs. 4(a) and 4(b) show the calculated one-particle 
densities of states at $T=T_{\rm c}$ (dashed lines) and $T=0.0030$ 
(solid lines) for $\delta=0.1451 (T_{\rm c}=0.0060)$ 
and $\delta=0.2037 (T_{\rm c}=0.0045)$, respectively. 
It is seen in Fig. 4 that the pseudo-gap opens at $T=T_{\rm c}$ 
due to the antiferromagnetic spin fluctuation. 
The superconducting gaps estimated from these figures are 
$2\Delta=0.0663$ for $\delta=0.1451 
(2\Delta/k_{\rm B}T_{\rm c}=11.0)$ 
and $2\Delta=0.0438$ for $\delta=0.2037 
(2\Delta/k_{\rm B}T_{\rm c}=9.7)$. 
The deviation of the frequency of the minimum density of state 
from the Fermi level at $T=0.0030$ for $\delta=0.2037$ seems to 
indicate that a size-effect still remains in the numerical work. 

\begin{figure}[ht]
\begin{center}
\resizebox{70mm}{!}{\includegraphics{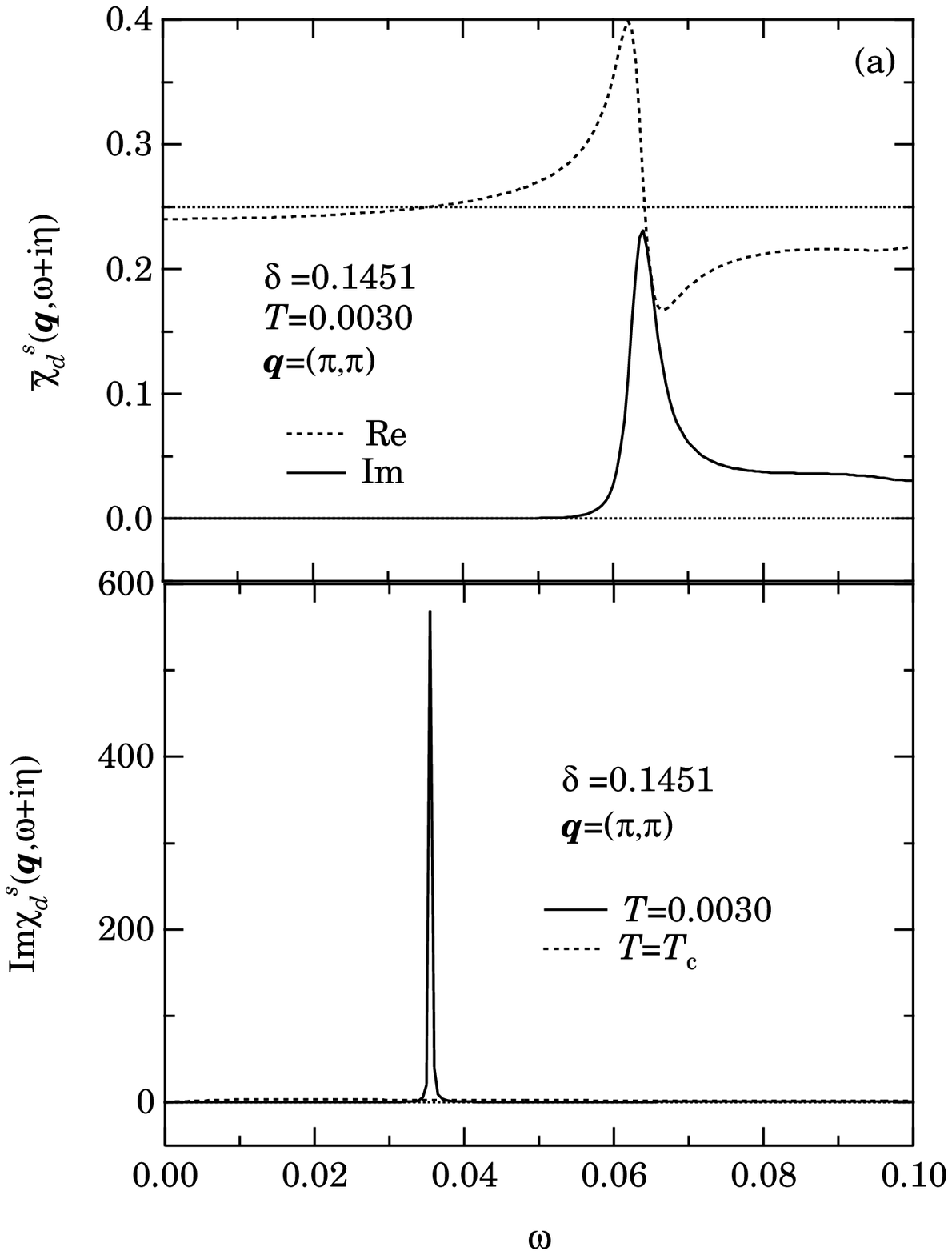}}\hspace*{5mm}
\resizebox{70mm}{!}{\includegraphics{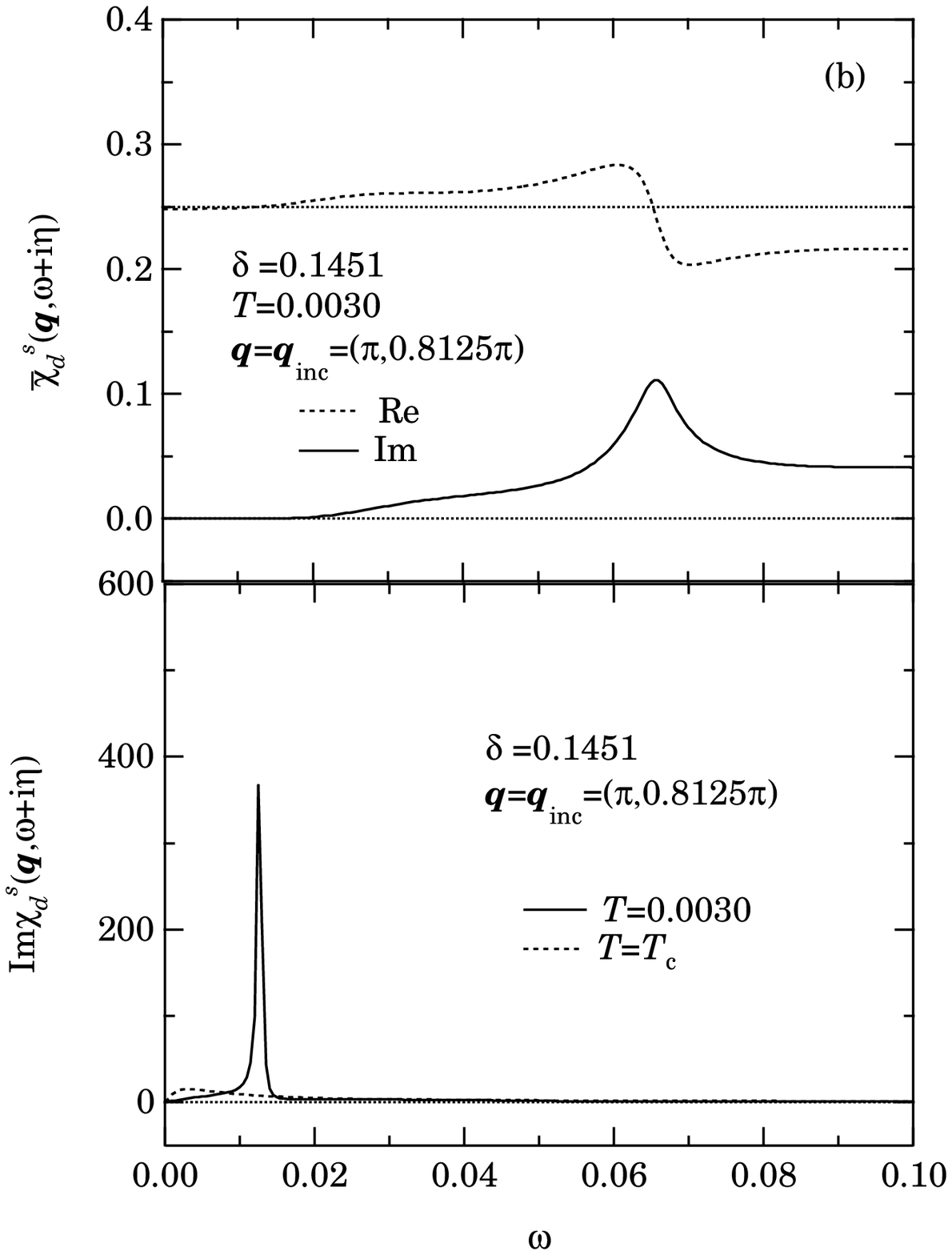}}
\end{center}
{\small 
\begin{flushleft}
Fig. 5. 
Calculated dynamical susceptibilities; upper panels show 
the real and imaginary parts of the irreducible susceptibilities 
and the lower panels show the final results for $\delta=0.1451$. 
(a) $\mib{q}=(\pi,\pi)$, (b) 
$\mib{q}=\mib{q}_{\rm inc}=(\pi,0.8125\pi)$.
\end{flushleft}}
\end{figure}

\begin{figure}[t]
\begin{center}
\resizebox{70mm}{!}{\includegraphics{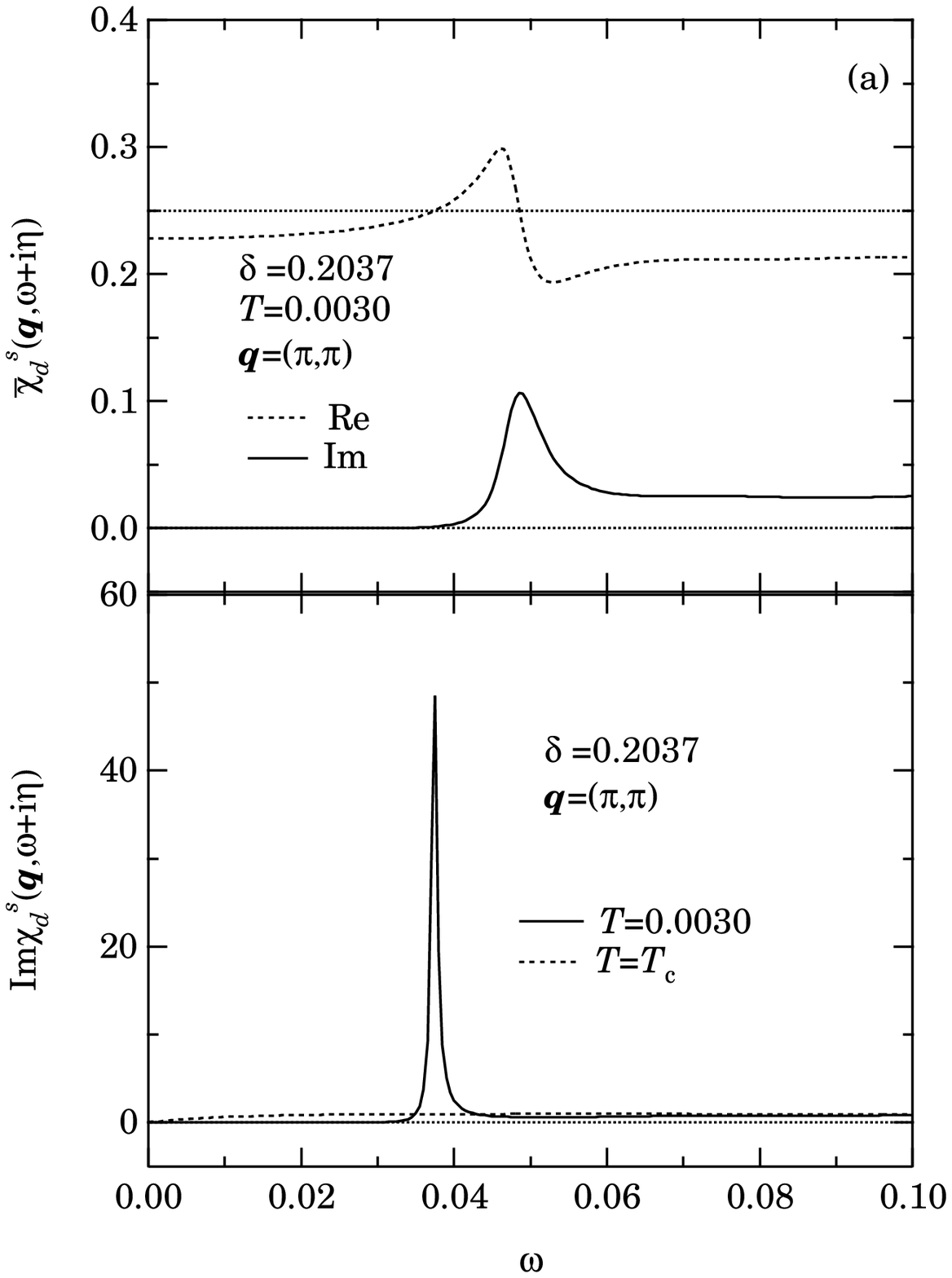}}\hspace*{5mm}
\resizebox{70mm}{!}{\includegraphics{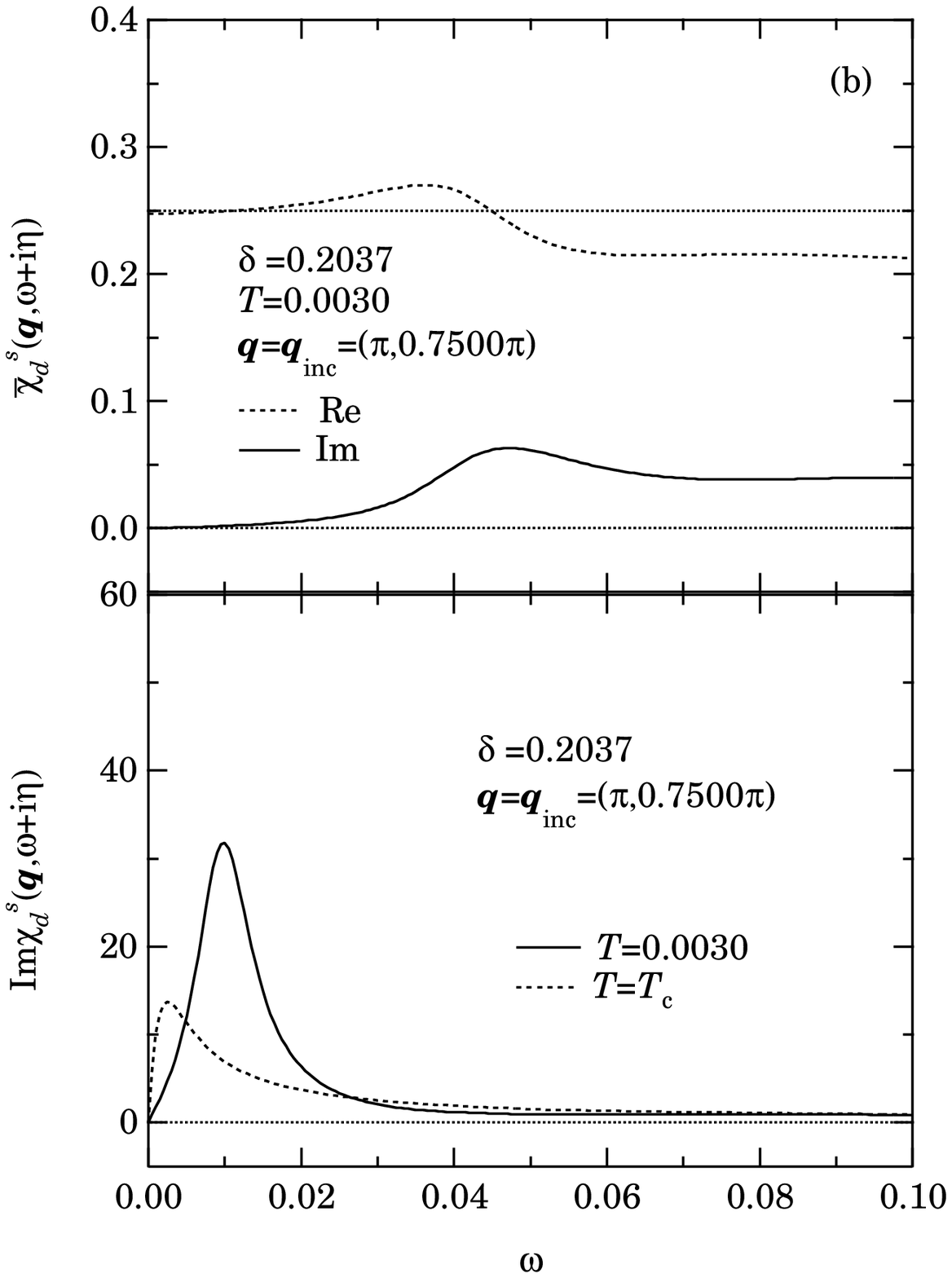}}
\end{center}
{\small 
\begin{flushleft}
Fig. 6. 
Calculated dynamical susceptibilities; upper panels show 
the real and imaginary parts of the irreducible susceptibilities 
and the lower panels show the final results for $\delta=0.2037$. 
(a) $\mib{q}=(\pi,\pi)$, (b) 
$\mib{q}=\mib{q}_{\rm inc}=(\pi,0.7500\pi)$.
\end{flushleft}}
\end{figure}

\subsection{Dynamical susceptibility.}
     We first show in Figs. 5(a) and 5(b) the $\omega$-dependence 
of the dynamical susceptibility in the superconducting state 
for $\delta=0.1451$ for the wave vectors 
$\mib q=\mib Q=(\pi,\pi)$ and $\mib q=\mib{q}_{\rm inc}$, 
respectively, where $\mib{q}_{\rm inc}$ is the wave vector 
for the incommensurate peak of the susceptibility. 
The upper panels show the real and imaginary parts of 
the irreducible susceptibility 
$\overline{\chi}_{d}^{\rm s}(\mib q,\omega+{\rm i}\eta)$ 
and the lower panels show the imaginary part of 
$\chi_{d}^{\rm s}(\mib q,\omega+{\rm i}\eta)$. 
The results of the same calculation for $\delta=0.2037$ 
are shown in Fig. 6. From these figures we find that 
${\rm Im}\overline{\chi}_{d}^{\rm s}(\mib q,\omega+{\rm i}\eta)$ 
has a peak at the value of $\omega=2\Delta_{\rm max}$ 
corresponding roughly to the energy gap in the density of states 
shown in Fig. 2: $2\Delta_{\rm max}=0.064$ for $\delta=0.1451$ 
and $2\Delta_{\rm max}=0.049$ for $\delta=0.2037$. On the other hand, 
${\rm Im}\chi_{d}^{\rm s}(\mib q,\omega+{\rm i}\eta)$ shows 
a sharp resonance peak at a frequency substatially smaller than 
$2\Delta_{\rm max}$. This peak is considered to be specific 
to the superconducting state since it tends to vanish at $T_{\rm c}$ 
as may be seen in Figs. 5 and 6. From the same figures we may see 
that the resonance peaks grow at the expense of the lower frequency 
spin fluctuations in the normal state. Also, the peak intensity 
seems to be stronger when the energy gap is larger. The resonance 
nature of this peak may best be seen from the following expression: 
\begin{equation}
  {\rm Im}\chi_{d}^{\rm s}(\mib q,\omega+{\rm i}\eta)=\frac
  {{\rm Im}\overline{\chi}_{d}^{\rm s}(\mib q,\omega+{\rm i}\eta)}
  {[1-U{\rm Re}\overline{\chi}_{d}^{\rm s}(\mib q,\omega+{\rm i}\eta)]
   ^{2}
  +[U{\rm Im}\overline{\chi}_{d}^{\rm s}(\mib q,\omega+{\rm i}\eta)]
  ^{2}}.
\end{equation}
and Figs. 5 and 6. In the denominator of the right hand side of 
this expression the second term may be quite small for the energies 
less than the superconducting gap energy. When the first term 
vanishes at $\omega=\omega_{\mibs q\mib{\mbox{}}}$, 
${\rm Im}\chi_{d}^{\rm s}(\mib q,\omega+{\rm i}\eta)$ naturally 
shows a sharp peak there. In Figs. 5 and 6 the resonance frequency 
$\omega_{\mibs q\mib{\mbox{}}}$ is given by the point where 
${\rm Re}\overline{\chi}_{d}^{\rm s}(\mib q,\omega+{\rm i}\eta)$ 
crosses the horizontal dotted lines indicating $1/U$. The peak width 
is given roughly by the magnitude of 
$\gamma_{\mibs q\mib{\mbox{}}}\equiv{\rm Im}
\overline{\chi}_{d}^{\rm s}(\mib q,
\omega_{\mibs q\mib{\mbox{}}}+{\rm i}\eta)$. Figs. 7(a) and 7(b) 
show the calculated $\mib q$-dependences of 
$\omega_{\mibs q\mib{\mbox{}}}$ and $\gamma_{\mibs q\mib{\mbox{}}}$ 
for $\delta=0.1451$ and $\delta=0.2037$, respectively. 
It is interesting to note that $\omega_{\mibs q\mib{\mbox{}}}$-
$\mib q$ curves are qualitalively similar to the corresponding 
results for the Hubbard model.\cite{rf:9} 

     Next we show in Figs. 8(a) and 8(b) the $\mib q$-integrated 
or local dynamical susceptibility ${\rm Im}\chi(\omega)$ 
for $\delta=0.1451$ and $\delta=0.2037$, respectively. 
Here the solid and dashed lines show the results at 
$T=0.0030$ and $T=T_{\rm c}$, respectively. 

     Finally we note that the resonance peaks calculated here and 
the result of calculation for ${\rm Im}\chi(\omega)$ bear 
striking resemblance to the results of neutron inelastic scattering 
experiments on ${\rm YBa}_{2}{\rm Cu}_{3}{\rm O}_{7}$ 
\cite{rf:19, rf:20, rf:21, rf:22} and 
${\rm YBa}_{2}{\rm Cu}_{3}{\rm O}_{6.5}$.\cite{rf:23} 
For more convincing comparisons, however, further experimental 
and theoretical studies are needed. 

\begin{figure}[ht]
\begin{center}
\resizebox{70mm}{!}{\includegraphics{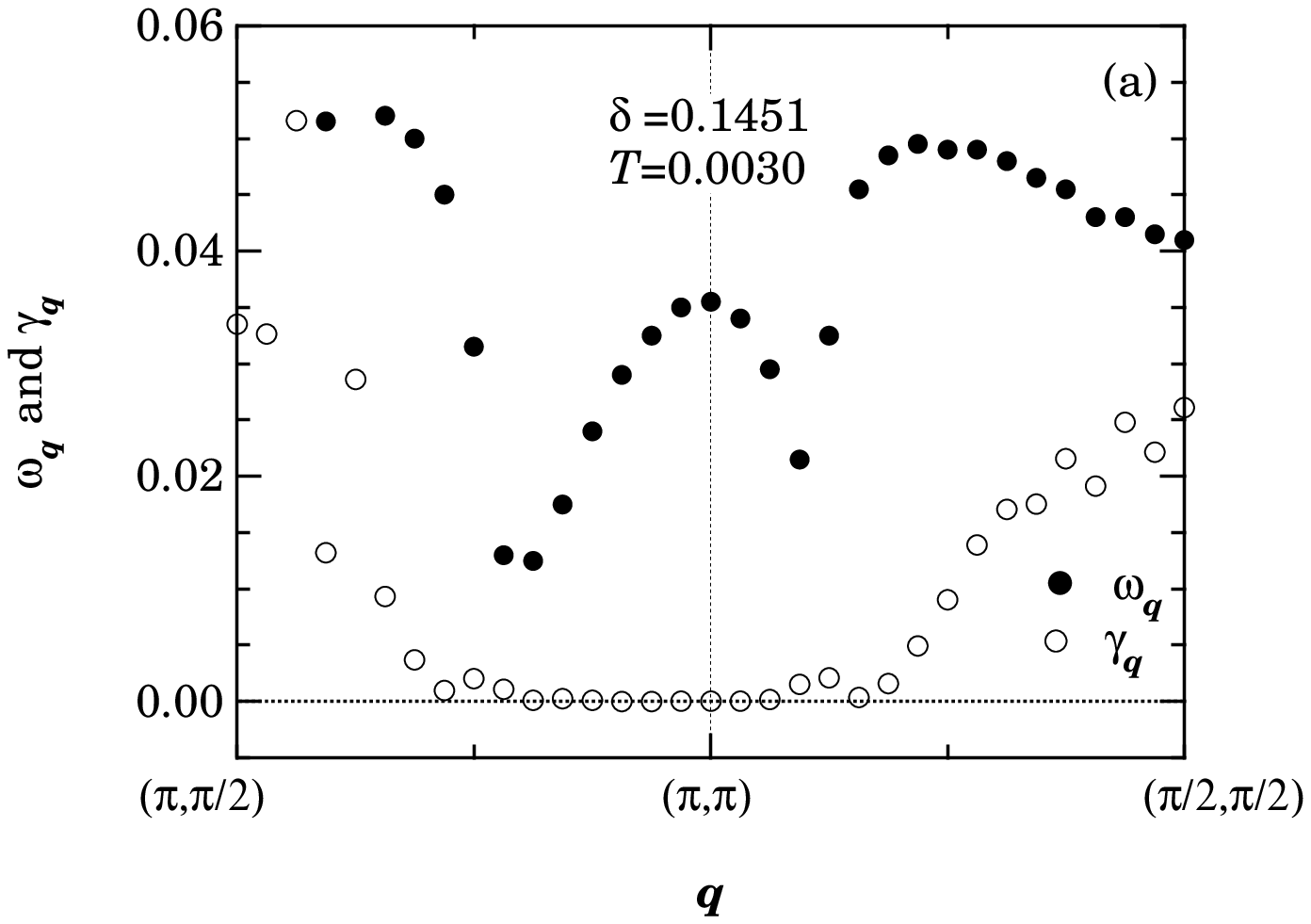}}\hspace*{5mm}
\resizebox{70mm}{!}{\includegraphics{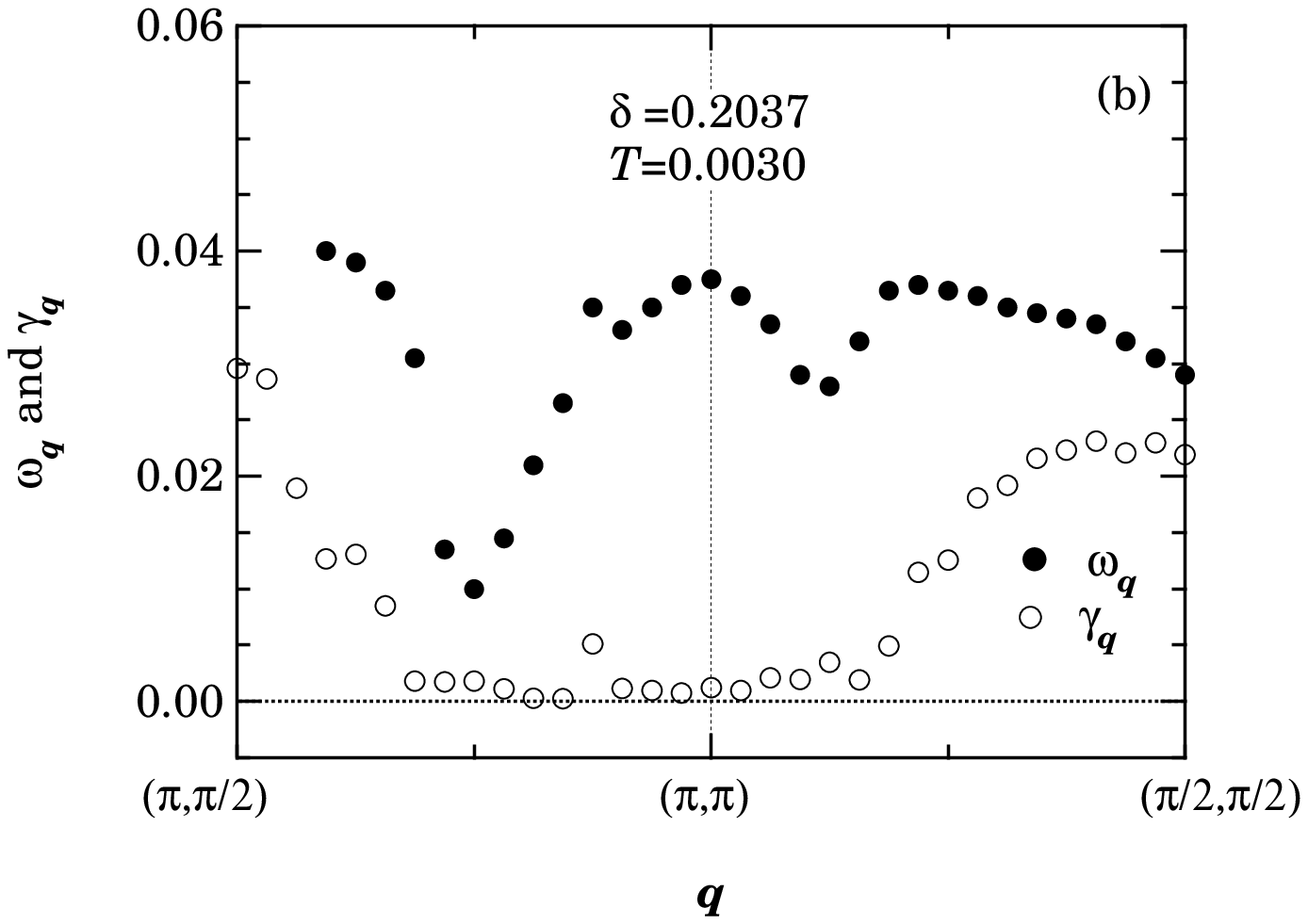}}
\end{center}
{\small 
\begin{flushleft}
Fig. 7. 
The resonance peak positions of the dynamical susceptibility 
and the damping constant. (a) $\delta=0.1451$, (b) $\delta=0.2037$. 
\end{flushleft}}
\end{figure}

\begin{figure}[ht]
\begin{center}
\resizebox{70mm}{!}{\includegraphics{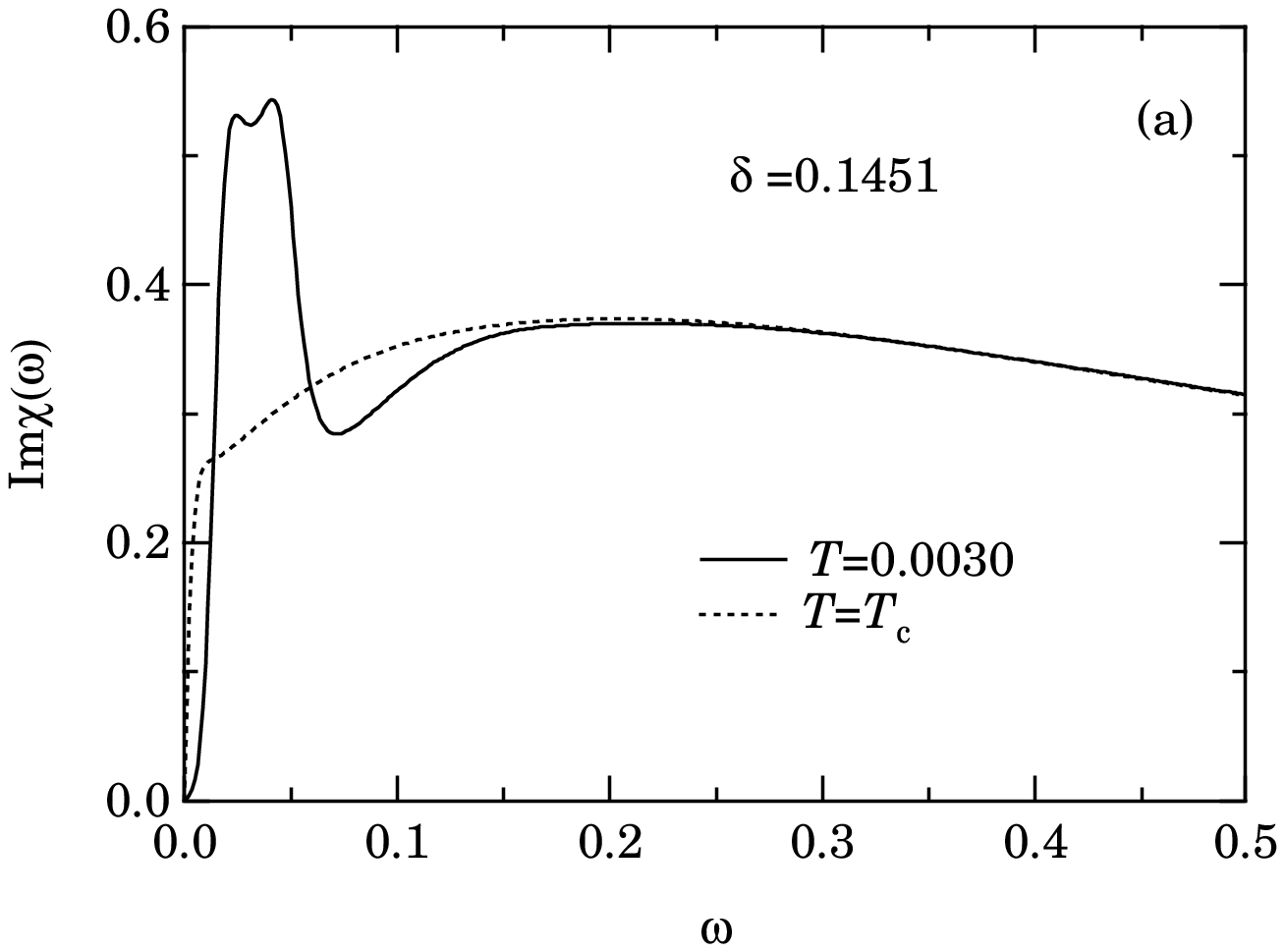}}\hspace*{5mm}
\resizebox{70mm}{!}{\includegraphics{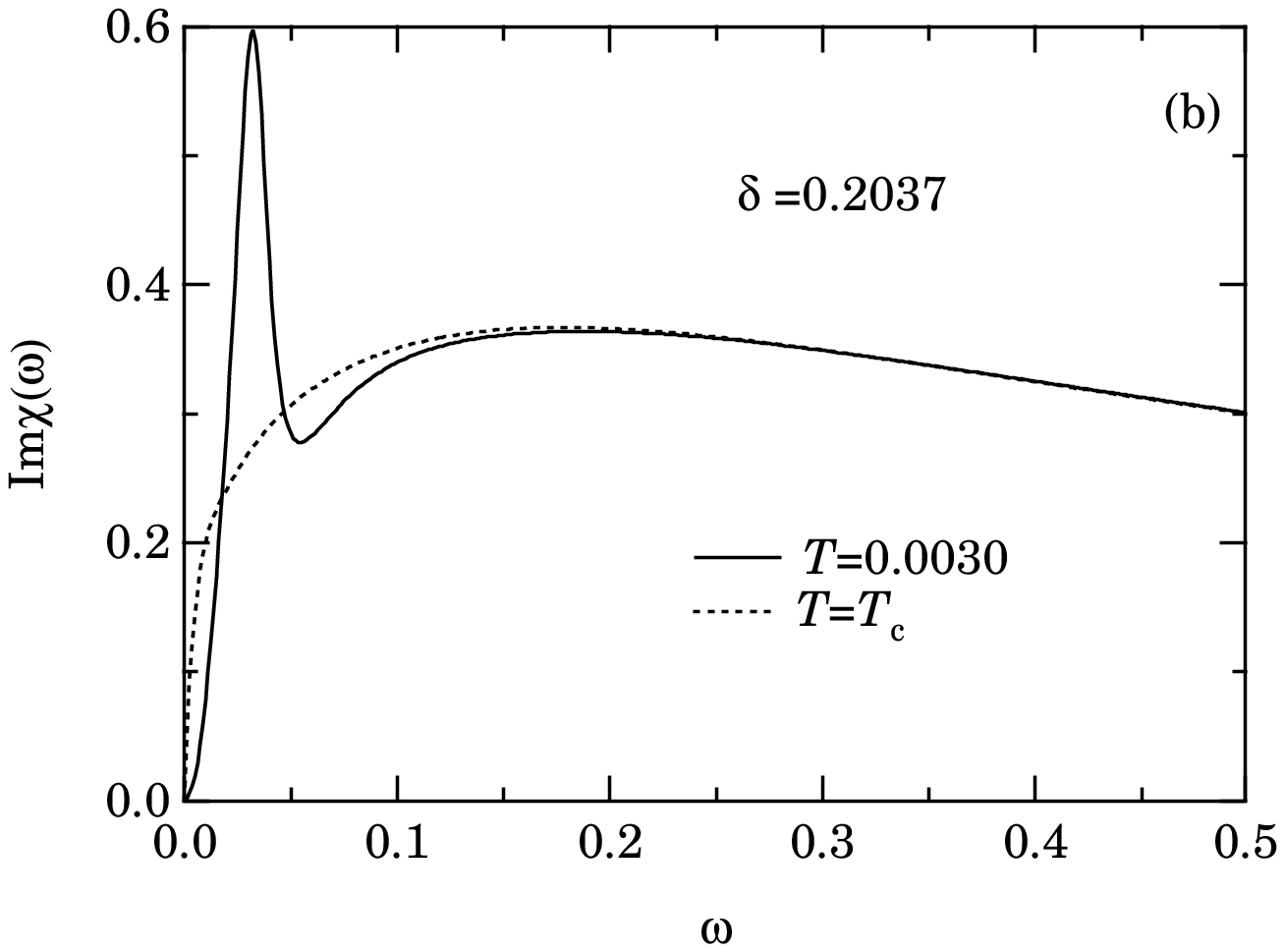}}
\end{center}
{\small 
\begin{center}
Fig. 8. 
Imaginary parts of the local or $\mib q$-integrated 
dynamical susceptibility. (a) $\delta=0.1451$, (b) $\delta=0.2037$. 
\end{center}}
\end{figure}

\begin{figure}[tp]
\begin{center}
\resizebox{70mm}{!}{\includegraphics{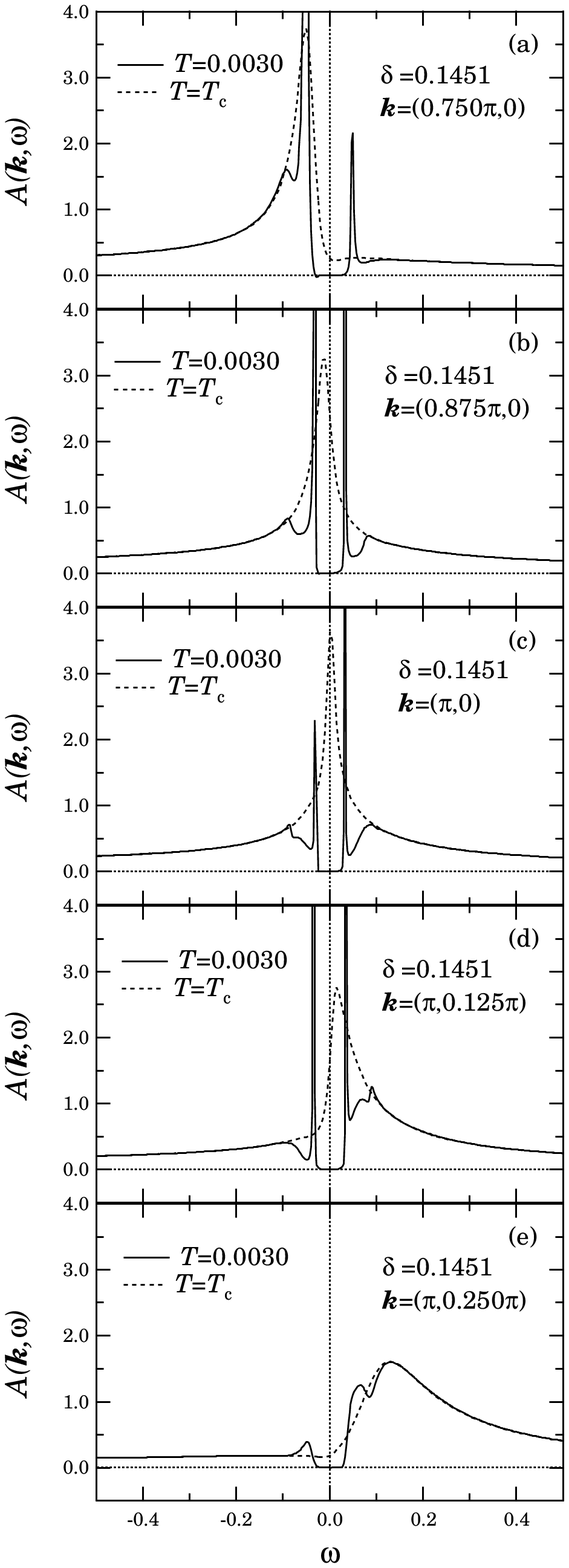}}\hspace*{5mm}
\resizebox{70mm}{!}{\includegraphics{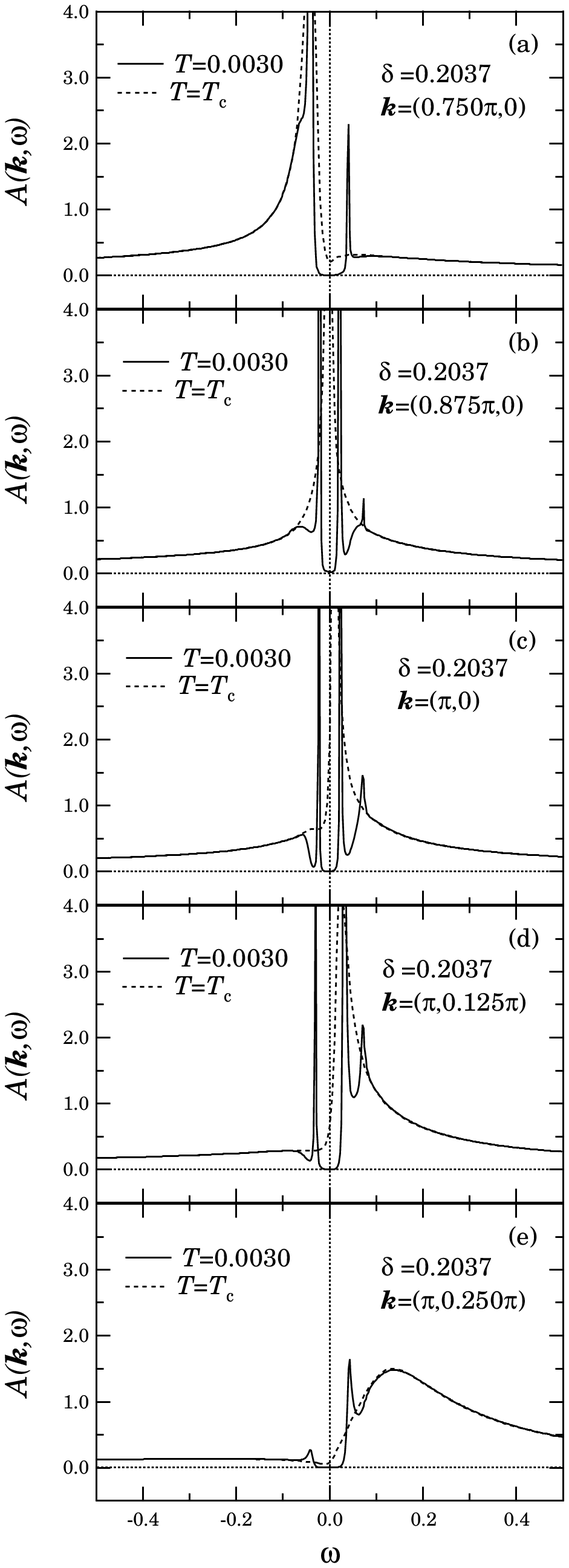}}
\end{center}
{\small 
\begin{center}
Fig. 9. 
One-electron spectral density around $\mib{k}=(\pi,0)$ 
for $\delta=0.1451$.\\
Fig. 10. 
One-electron spectral density around $\mib{k}=(\pi,0)$ 
for $\delta=0.2037$.
\end{center}}
\end{figure}

\begin{figure}[ht]
\begin{center}
\resizebox{70mm}{!}{\includegraphics{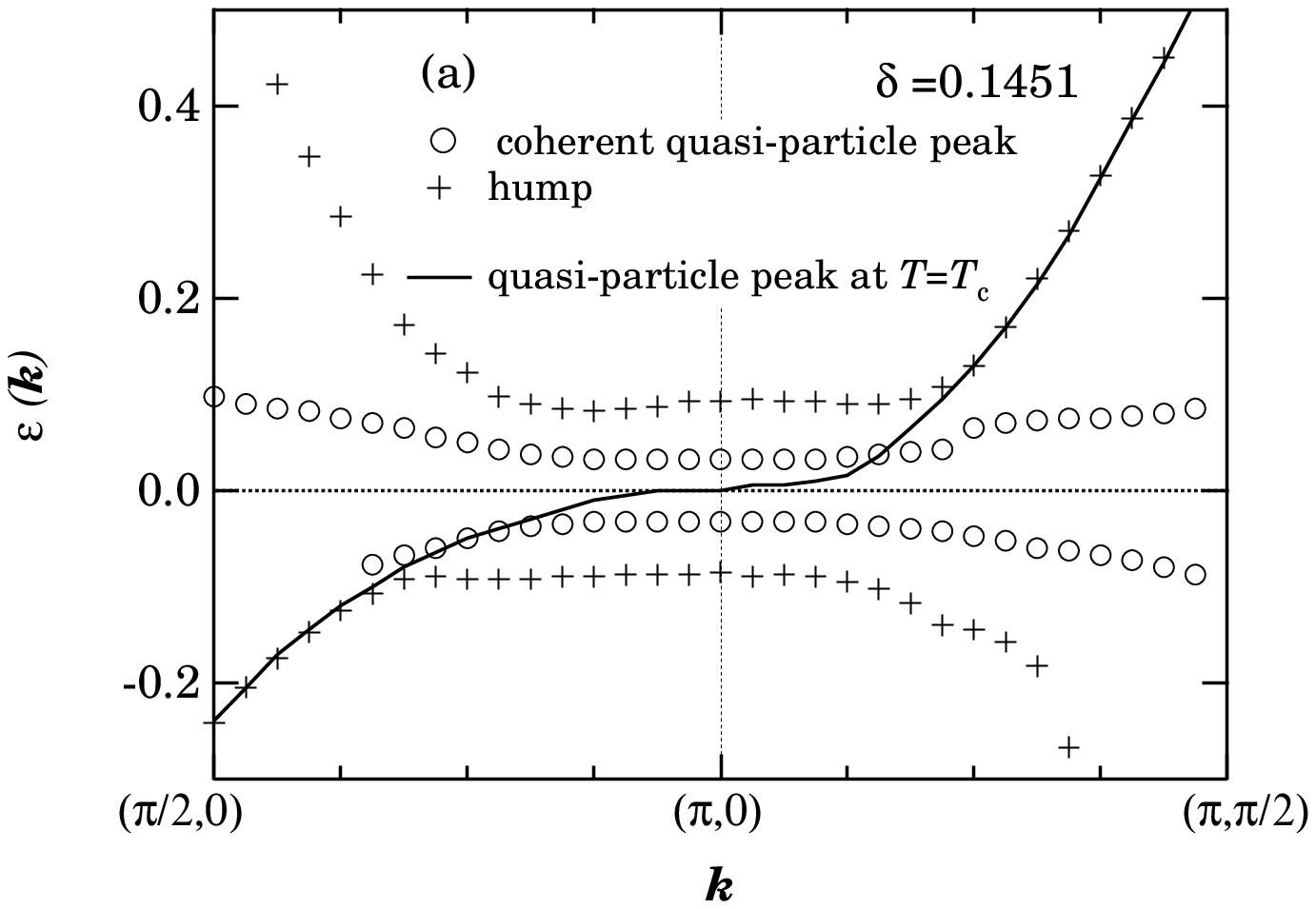}}\hspace*{5mm}
\resizebox{70mm}{!}{\includegraphics{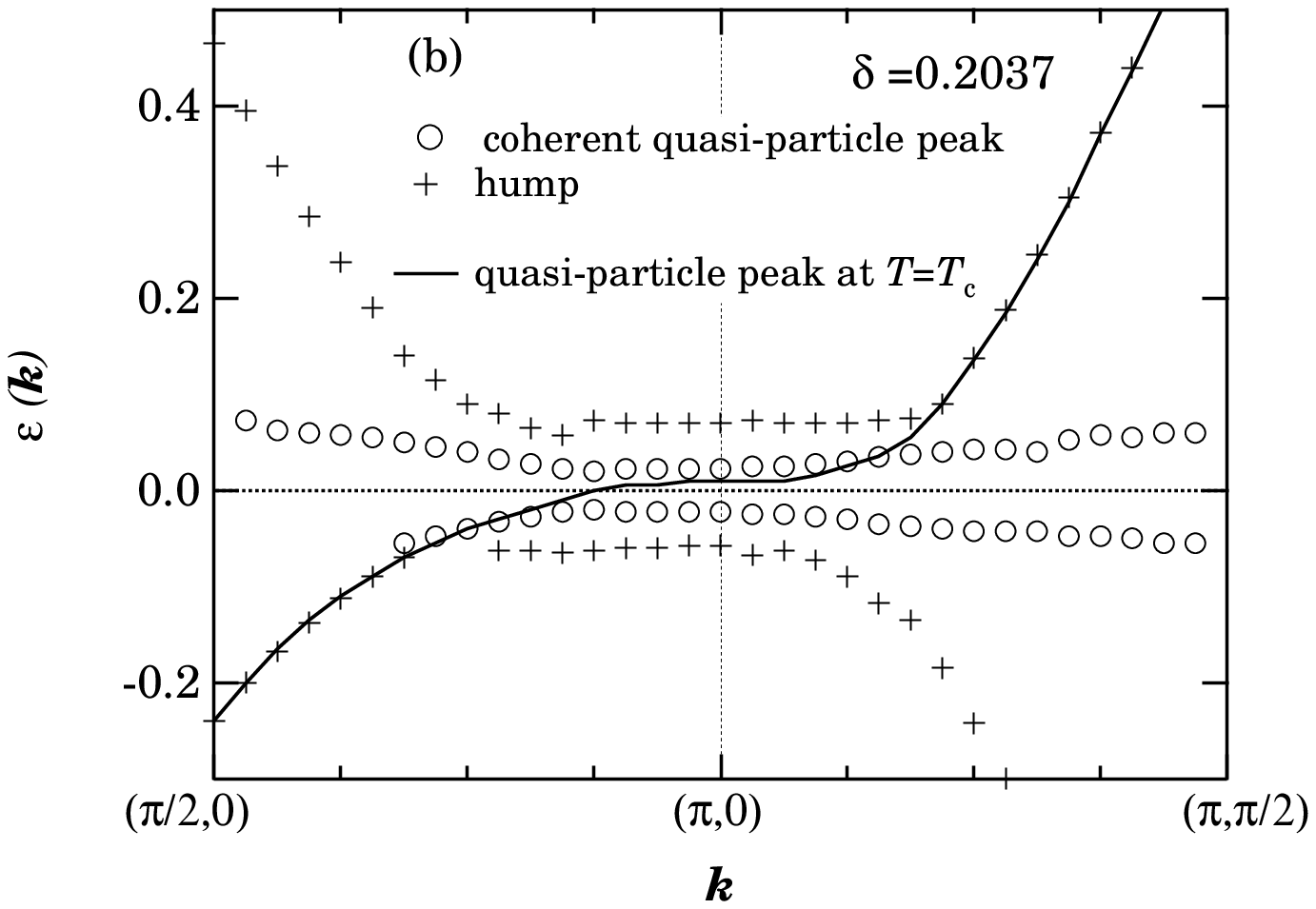}}
\end{center}
{\small 
\begin{flushleft}
Fig. 11
Energy dispersions of the coherent quasi-particle peak 
and hump in the superconducting state ($T=0.0030$). 
(a) $\delta=0.1451$, (b) $\delta=0.2037$.
\end{flushleft}}
\end{figure}

\subsection{One electron spectral density.}
     We show in Fig. 9 the calculated one-particle spectral density 
$A(\mib k,\omega)$ for $\delta=0.1451$ 
for five wave vectors around $\mib k=(\pi,0)$, i.e.,
(a) $\mib k=(0.750\pi,0)$, 
(b) $\mib k=(0.875\pi,0)$, 
(c) $\mib k=(\pi,0)$, 
(d) $\mib k=(\pi,0.125\pi)$ and 
(e) $\mib k=(\pi,0.250\pi)$, where the solid and 
dashed lines show the results for $T=0.0030$ and $T=T_{\rm c}$, 
respectively. 
Corresponding results for $\delta=0.2037$ are shown in Fig. 10. 
From Figs. 9 and 10 we find a remarkable feature of 
$A(\mib k,\omega)$ in the superconducting state, 
consisting of a sharp coherent quasi-particle peak 
followed by a structure of dip and hump. 
As a matter of fact 
these behaviors are quite similar to those observed in 
the angle-resolved photoemission spectrum.
\cite{rf:24, rf:25, rf:26, rf:27} 

     Next we show in Figs. 11(a) and 11(b) the dispersions of 
the coherent quasi-particle peaks and the humps together with 
the quasi-particle dispersion at $T=T_{\rm c}$ 
around $\mib k=(\pi,0)$ for $\delta=0.1451$ and $\delta=0.2037$, 
respectively. The dispersions are defined by the value of 
$\epsilon(\mib k)$ giving a maximum value of $A(\mib k,\omega)$ 
for a given $\mib k$. These figures show that the quasi-particle 
peaks are almost dispersionless while the humps become 
indistinguishable from those in the normal state as 
we go away from the Fermi level. This behavior is also consistent 
with the results of ARPES measurements.\cite{rf:27} 

     In order to get better understanding of the above mentioned 
specific feature of the one-particle spectral density 
in the superconducting state, we show in Fig. 12 the calculated 
self-energy and the one-particle spectral density of the $d$-state, 
since the spectral density is dominated by this contribution. 
We have
\begin{eqnarray}
  &&\frac{1}{G_{dd}({\mib k},\omega+{\rm i}\eta)}
 \equiv\frac{1}{G_{dd}^{(0)}({\mib k},\omega+{\rm i}\eta)}
 -\Sigma(\mib k,\omega+{\rm i}\eta),\\
 &&A_{d}(\mib k,\omega)\equiv-\frac{1}{\pi}{\rm Im}
 G_{dd}({\mib k},\omega+{\rm i}\eta).
\end{eqnarray}
\begin{eqnarray}
  &&\Sigma(\mib k,\omega+{\rm i}\eta)
  =\Sigma^{(1)}(\mib k,\omega+{\rm i}\eta)
  -G_{\rm N}^{*}({\mib k},-\omega+{\rm i}\eta)
   [\Sigma^{(2)}(\mib k,\omega+{\rm i}\eta)]^{2},\\
  &&\nonumber\\
  &&\frac{1}{G_{\rm N}({\mib k},\omega+{\rm i}\eta)}
 \equiv\frac{1}{G_{dd}^{(0)}({\mib k},\omega+{\rm i}\eta)}
 -\Sigma^{(1)}(\mib k,\omega+{\rm i}\eta).
\end{eqnarray}
A quasi-particle peak is expected at the frequency which satisfies 
\begin{equation}
  \frac{1}{G_{dd}^{(0)}({\mib k},\omega+{\rm i}\eta)}
 -{\rm Re}\Sigma(\mib k,\omega+{\rm i}\eta)=0,
\end{equation}
provided ${\rm Im}\Sigma(\mib k,\omega+{\rm i}\eta)$ is small 
at the same time. As is seen in Fig. 12 the superconducting 
energy gap gives almost vanishing values of 
${\rm Im}\Sigma(\mib k,\omega+{\rm i}\eta)$ in a fairly wide 
frequency range around $\mib{k}=(\pi,0)$ and sharp quasi-particle 
peaks are expected in this region of $\mib k$-space. 
As the frequency increases beyond this region 
$|{\rm Im}\Sigma(\mib k,\omega+{\rm i}\eta)|$ increases first 
rapidly and then gradually. At the same time the value of 
$[G_{dd}^{(0)}({\mib k},\omega+{\rm i}\eta)]^{-1}
 -{\rm Re}\Sigma(\mib k,\omega+{\rm i}\eta)$
increases rapidly first to form a dip structure in 
$A_{d}(\mib k,\omega)$ and then the $\omega$-dependence of this 
value reflect itself on the spectrum of $A_{d}(\mib k,\omega)$ 
as is expected from eqs. (27, 28) and is actually seen in Fig. 12.

\begin{figure}[ht]
\begin{center}
\resizebox{168mm}{!}{\includegraphics{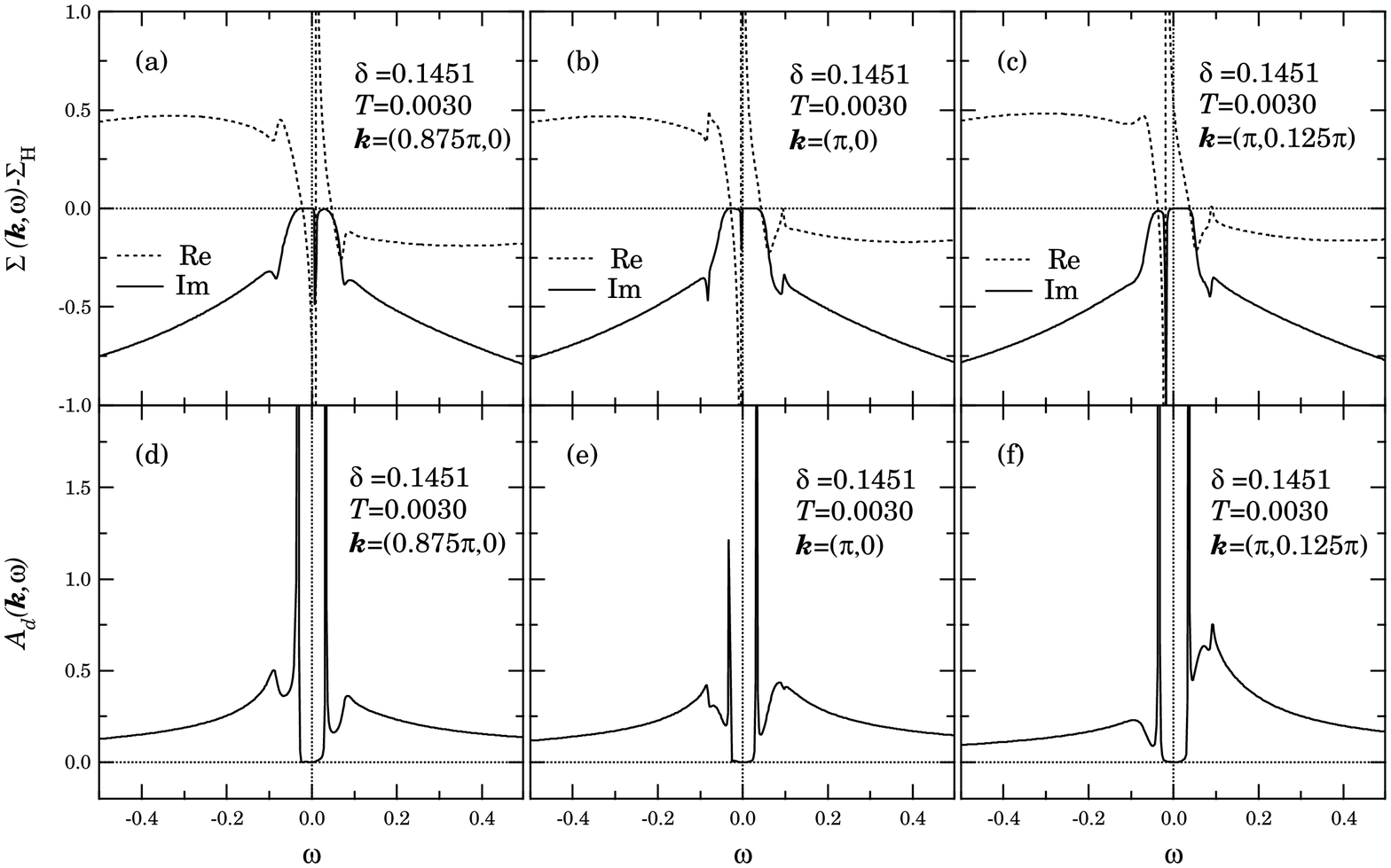}}
\end{center}
{\small 
\begin{flushleft}
Fig. 12. 
Self-energy of the $d$-component of the Green's function 
and its one-particle spectral density at three points in the 
$\mib k$-space around $\mib{k}=(\pi,0)$.
\end{flushleft}}
\end{figure}

\section{Conclusion and Discussion}
     On the basis of the $d$-$p$ model we have studied the 
unconventional superconducting state induced by the 
antiferromagnetic spin fluctuations in the optimal and over-doped 
regimes of high-$T_{\rm c}$ cuprates. 
Calculations were performed 
for the superconducting oder parameter of $d_{x^2 - y^2}$ symmetry, 
the dynamical susceptibility and the one-electron spectral density, 
by using the FLEX approximation, i.e., the strong coupling theory 
with the self-consistently renormalized random phase approximation 
(RRPA) for spin fluctuations. 
The order parameter or 
the gap function was found to develop more rapidly than 
in the BCS model, and the ratio of the low temperature maximum 
gap energy to $k_{\rm B}T_{\rm c}$ was about 10. 
The calculated spin fluctuation spectrum below $T_{\rm c}$ 
showed a resonance peak around $\mib{q}=(\pi,\pi)$ 
whose intensity was especially strong around the optimal doping 
concentration and its behavior bears striking resemblance to 
the peak at 41 meV observed in the neutron scattering experiment 
on ${\rm YBa}_{2}{\rm Cu}_{3}{\rm O}_{7}$. 
The calculated one-particle spectral density showed a sharp 
quasi-particle peak around $\mib{k}=(\pi,0)$ followed by 
a structure with a dip and a hump. These results were quite 
similar to those observed in the ARPES spectra. 

     In view of these results we may conclude that the present 
calculation gives an additional support to the spin fluctuation 
mechanism for the high-$T_{\rm c}$ cuprates, especially in the 
optimal and over-doped concentration regimes. 

     Desired improvements of the present approach include the 
consideration of the vertex corrections neglected in the FLEX 
approximation. Although the effect was suggested to be 
insignificant or rather favorable for the superconductivity,
\cite{rf:28} a self-consistent calculation with the vertex 
corrections is desired. The effect of impurity potential is 
expected to be more significant in the superconducting state 
than in the normal state. For example, the observed linear 
temperature dependences of $1/T_{1}$ and the specific heat 
far below $T_{\rm c}$ were successfully interpreted in terms of 
the additional density of states around the Fermi level 
caused by the impurity effect.\cite{rf:29} 
This effect will also have significant influence on 
the resonance peak in the dynamical susceptibility and on 
the above-discussed structures in the one-electron spectral density. 

     Of course there are many problems still to be clarified, 
particularly in the under-doped regime. The most controversial 
among them may be the pseudo-gap phenomena in the under-doped 
regime observed in NMR,\cite{rf:30} ARPES,
\cite{rf:31,rf:32,rf:33,rf:34,rf:35} and other measurements. 
In the present approach we treat the spin fluctuations 
in the spirit of the self-consistent renormalization (SCR) theory 
which applies around the antiferromagnetic instability. 
However, the FLEX approximation without the vertex corrections 
may not be sufficient particularly in dealing with the low 
frequency phenomena. Also, it seems possible that the ground state 
in the under-doped regime has an antiferromagnetic long range 
order extending to the region rather far from the instability point. 
If such circumstances are really the case it seems necessary 
to develop an approximation beyond the conventional SCR, 
taking account of substantially larger effects of non-linear 
mode-mode couplings.

\section*{Acknowledgements}
We are indebted Dr. S. Nakamura and Dr. H. Kondo for useful 
discussions particularly on numerical calculations.

\end{document}